\RequirePackage{fix-cm}
\documentclass[twocolumn]{svjour3}          
\smartqed  
\usepackage[utf8]{inputenc} 
\usepackage{graphicx}

\usepackage{multirow}
\usepackage[normalem]{ulem}
\useunder{\uline}{\ul}{}
\usepackage{longtable}
\usepackage[square,sort,comma,numbers]{natbib}
\usepackage{caption}
\UseRawInputEncoding 

\begin{document}

\title{ A Bibliography of Multiple Sclerosis Lesions Detection Methods using Brain MRIs
}





\author{Atif Shah$^{1,2}$ \and
        Maged S. Al-Shaibani$^2$ \and
        Moataz Ahmad$^2$ \and
        Reem Bunyan$^3$
}


\institute{ Atif Shah \at {atif.shah@oulu.fi}, {atifshahcs@gmail.com} \\ \\
             1. \emph{Center for Machine Vision and Signal Analysis (CMVS), University of Oulu, Oulu, Finland.}  
           \and
            \at
             2. Information and Computer Science Department, KFUPM, Dhahran 31261, Saudi Arabia\\
          \and
           \at
            3. Information and Computer Science Department, KFUPM, Dhahran 31261, Saudi Arabia\\
         \and
          \at
           4. Neurosciences Center, King Fahd Specialist Hospital Dammam, Dammam, Saudi Arabia 
}

\date{Received: date / Accepted: date}

\maketitle

\begin{sloppypar}
\begin{abstract}
\textit{Introduction:} Multiple Sclerosis (MS) is a chronic disease that affects millions of people across the globe.  MS can critically affect different organs of the central nervous system such as the eyes, the spinal cord, and the brain. 

\textit{Background:} To help physicians in diagnosing MS lesions, computer-aided methods are widely used. In this regard, a considerable research has been carried out in the area of automatic detection and segmentation of MS lesions in magnetic resonance images (MRIs). 

\textit{Methodology:} In this study, we review the different approaches that have been used in computer-aided detection and segmentation of MS lesions.  Our review resulted in categorizing MS lesion segmentation approaches into six broad categories: data-driven, statistical, supervised machine learning, unsupervised machine learning, fuzzy, and deep learning-based techniques. We critically analyze the different techniques under these approaches and highlight their strengths and weaknesses. 

\textit{Results:} From the study, we observe that a considerable amount of work, around 25\% of related literature, is focused on statistical-based MS lesion segmentation techniques, followed by 21.15\% for data-driven based methods, 19.23\% for deep learning and 15.38\% for supervised methods. 

\textit{Implication:} The study points out the challenges/gaps to be addressed in future research. The study shows the work which has been done in last one decade in detection and segmentation of MS lesions. The results show that, in recent years, deep learning methods are outperforming all the others methods.

\end{abstract}

\section{Introduction}
\label{intro}
\begin{figure*}
    \centering
    \includegraphics[width=1.0\textwidth]{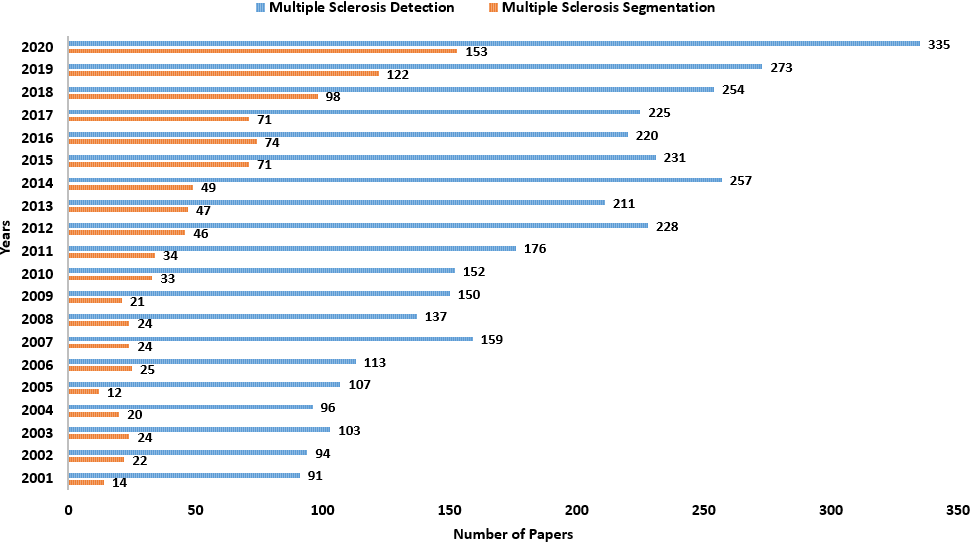}
    \caption{MS papers statistics for period of 2001 - 2020}
    \label{fig:stats}
\end{figure*}

Multiple Sclerosis (MS) is a progressive disease that mostly affects young adults between the ages of 20 and 40 years. 10\% of these patients feel healthy for 20 years; however, 50\% of patients need help walking for up to 15 years and in some cases, they die within a few months. Studies show a correlation between the age of the patient and the severity of the disease; the younger the patient, the more severe the disease is \citep{noseworthy2000a}. According to the Multiple Sclerosis Foundation, 400,000 in the United States and 2.5 million people around the world have MS. The ratio of MS in women to men is 2:1 \citep{multiple_sclerosis_foundation}. In the United States, the treatment of MS is expensive with yearly costs ranging from \$8,528 to \$52,244 per patient \citep{adelman2013a}. There is no single test that would suffice by itself to determine a diagnosis. Still, the patient history, along with a neurological examination and a collection of tests which may include blood tests, spinal fluid analysis, and MRIs, could help. The exact cause of MS is still unknown, resulting in a no known prescription.
MS is a chronic disease of the Central Neural System (CNS) affecting the brain changing its structure and morphology and the spinal cord causing a lesion in the white matter of the brain by damaging myelin sheath \citep{roy2013a}. Depending on the regions affected, MS may affect different CNS organs such as the eyes, the spinal cord and the brain leading to blurry vision and muscle weakness. Atrophy of the spinal cord and the brain is the main part of MS pathology which is relevant to the progression of the disease \citep{roy2013a}. CNS atrophy, which consists of White Matter (WM) or Gray Matter (GM), becomes worse with the increase of the disease causing progressive loss of brain tissues with a rate of 0.6\% to 1.2\% per year. Brain atrophy takes place in cortical and sub-cortical regions even though it’s unknown if the atrophy arises in WM or GM. To estimate the overall GM damage, one can measure the GM volume which can be done via a conventional MRI \citep{horsfield2003a}. Axonal loss is the reason for atrophy which is correlated with the number of T2 lesions. In short clinical trials, the cerebral atrophy rate is also detectable in Relapsing-Remitting MS (RRMS) \citep{elskamp2010a}. Fisher et al\citep{fisher2008a}, Horakova et al \citep{horakova2008a} and Pagani et al \citep{pagani2005a} agreed that in GM regions the tissue loss is higher than WM, also GM atrophy is found in some patients very early while the WM atrophy varies in different RRMS stages. 
There have been numerous approaches to allow MS lesions segmentation in MRIs for the sake of learning more about the MS disease of a patient. However, to the best of our knowledge, there is no framework that can be used to evaluate the current state of affairs regarding the effectiveness of such approaches as well as identifying gaps for future research. In this paper, we present such a framework along with evaluating the current literature accordingly. We identified segmentation methods through research databases, which includes, IEEE Xplore, PubMed, Google Scholar and Scopus between January 2005 and December 2020. We considered the literature from the year 2005 since there are already some other studies that adequately discussed various methods before 2005 along with few other articles published after 2005 \citep{mortazavi2012a}, \citep{llad2012a}, \citep{caligiuri2015a}. There are many other recent studies to detect MS lesion via histological classification based on microscopic images \citep{kuhlmann2017a}.  Detection of veins in white matter lesion (WML) also helps in MS diagnosis \citep{campion2017a}. A Matrix-Assisted Laser Desorption Ionization Mass Spectrometry Imaging  has also been proposed to detect protein and peptides in tissues to help in MS diagnosis \citep{maccarrone2017a}. However, in this study, we focus on MS Computer-Aided Diagnosis (CAD) systems that use MRI brain images. In recent years, the number of MS CAD publications increases significantly as shown in Figure \ref{fig:stats}.

A recently published review paper by Danelakis et al \citep{danelakis2018a} has, similarly, considered evaluating the various MS lesion segmentation methods. However, our work is different as we categorized the methods into six different categories with more focus on deep learning techniques, while their study used only two generic categories of supervised and unsupervised. We included more than 24 recent papers focused on MS segmentation and detection via brain MRIs. Moreover, while Danelakis et al's study only focused only on segmentation methods, our study also considers the detection methods as well. Additionally, while their study only considered Dice Similarity Coefficient (DSC) as metric for evaluation, we consider all the metrics which are desired in CAD systems, i.e., Specificity, Sensitivity, Precision, True and False positive rates and DSC. We also did the analysis year-wise, method-wise, and also discussed future directions for detection and segmentation methods. Lastly, this study conducted more papers from previous studies as shown in Table \ref{tab:rev_papers} and also more focused on deep learning methods.

To summarize the contribution of this paper and its distinction compared to other reviews in the field, this paper; to the best of our knowledge, is the first review that proposes a detailed breakdown and categorization of the methods and techniques considered for the problem. We category the methods and techniques into six main categories with underneath subcategories. We extensively review more studies than any review we aware exists in the literature by a wide margin, as highlighted by Table \ref{tab:rev_papers}. Moreover, we consider more than 30 studies published since 2018 embracing the latest advances in deep learning methods and architectures.

The rest of the paper is organized as follows. Section \ref{sec:2} provides background on radiology images used in MS lesion detection and MS lesions types. Section \ref{sec:3}  discusses the overall flow of the CAD system. Section \ref{sec:4} discusses the datasets used in literature. Section \ref{sec:5} provides details of various evaluation metrics that have been used in literature. Section \ref{sec:6} presents our review methodology. Section \ref{sec:7} discusses various categories of MS lesion segmentation and detection approaches. Finally, Section \ref{sec:8} discusses challenges and future directions.

\begin{table}[]
\begin{tabular}{|l|l|l|}

\hline
Article &  Year & Number of paper reviewed \\ \hline
 Mortazavi et al \citep{mortazavi2012segmentation} & 2012 &  44\\ \hline
 Llad{\'o}\citep{llado2012automated} (a) & 2012 & 34 \\ \hline
 Llad{\'o} \citep{llado2012segmentation} (b) & 2012 & 34 \\ \hline
 Garcia et al \citep{garcia2013review} & 2013 & 55 \\ \hline
Danelakis et al \citep{danelakis2018a} & 2018 & 45 \\ \hline
 Current paper & 2021 & 156 \\ \hline
\end{tabular}
\caption{Number of papers reviewed in survey papers}
\label{tab:rev_papers}
\end{table}

\section{Background}
\label{sec:2}

\subsection{Imaging of MS Lesions}
\label{sec_sub:imaging}
\begin{figure}
    \centering
    \includegraphics[width=0.4\textwidth]{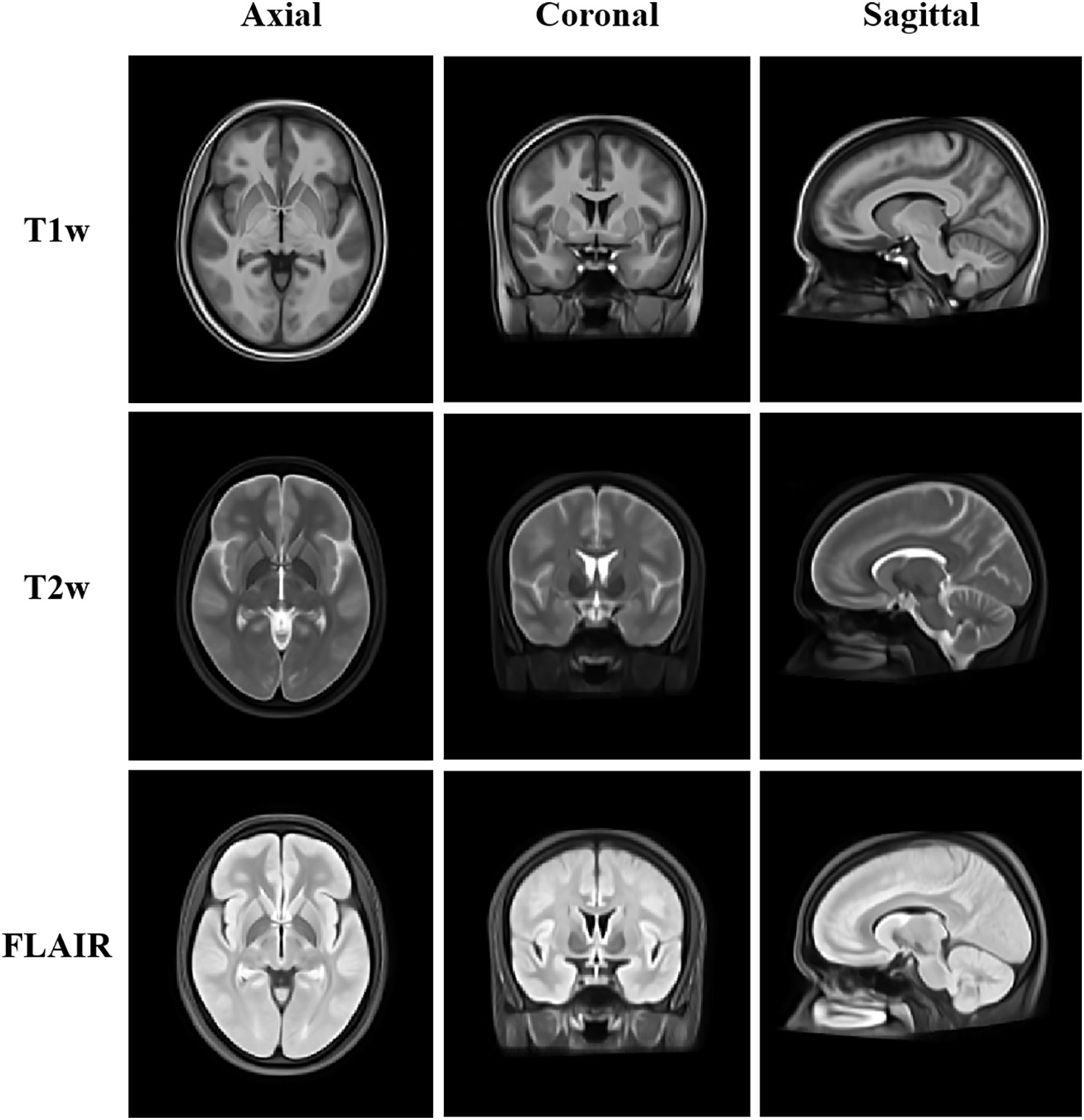}
    \caption{Different types of MRI's}
    \label{fig:mri}
\end{figure}

Medical imaging has been instrumental in detecting diseases such as MS. For example, different medical imaging techniques have been reported to play a vital role in detecting MS lesion in the brain; this includes Computed Tomography (CT) images, X-rays, ultrasound, and Magnetic Resonance Imaging (MRI). For instance, MRIs, linked with clinical judgment, have been used to provide essential information regarding the MS disease progress \citep{gonz2017a}\citep{breckwoldt2017a}\citep{nelson2017a}\citep{bishop2017a}. Experts use MRIs to find out effective therapeutic regimes. Based on sharp contrasts and much information, the MRI is superior to other images to study CNS diseases. MRI is considered the best preclinical examination for MS which can expose abnormality of 95\% of patients \citep{khayati2008a}. The traditional analysis of MRI is a difficult and time-consuming task as the results directly depend on the neurologist experience. The difficulty is related to the complexity of edge boundaries. To overcome these complexities, automatic segmentation is used as an important task to provide satisfactory performance.
Most of the medical research is focused on MRIs due to their better resolution and visualization of the brain with a different gray and white contrast. MRIs show the loss of myelin with hypointensities or hyperintensities with respect to its surrounding tissues. Various protocols are used to enhance the performance of the early and advanced stages of the disease. These protocols include T1- weighted (T1-w), the longitudinal relaxation time of tissue, T2-weighted (T2-w), the transverse relaxation time of tissue; Proton density-weighted (PD-w) and fluid-attenuated inversion recovery (FLAIR) brain images. PD-w or PD images are also used which has sharp contrast and high density of protons. In T1 images, the lesion consists of hypo-intensities, also known as black holes, while in T2, it consists of hyper-intensities, as shown in Figure \ref{fig:mri}. In FLAIR, the contrast of Cerebrospinal Fluid (CSF) and lesions are more highlighted. However, some lesions in T2 are similar to CSF in FLAIR. Most researcher prefers Axial FLAIR over Sagittal FLAIR shown in Figure \ref{fig:mri}  because Axial can expose more MS lesions. Galler et al \citep{galler2016a} compared these two types of scans and the results show that Axial can perform better and showed 22\% more lesions than Sagittal scans. Some researchers also used gadolinium \citep{cromb2015a}, also called MRI contrast media. These chemical substances are injected into the body before the MRI scans to enhance the contrast which can improve lesion segmentation. Gonyea et al \citep{gonyea2015a} used whole-brain T1-rho and T2 images to detect the lesion. T1-rho is used in musculoskeletal imaging which is not popular for clinical use. The lesion was manually segmented by an expert who shows T1-rho exposes more lesions than T2 images. Diffusion tensor and magnetization transfer are also MR imaging which can validate microstructural tissue abnormalities in MS patients \citep{filippi2017a}.

Some regions cannot be classified due to complicated brain tissues and the finite resolution of imaging which affects the voxels on the border of tissues. These border voxels are known as partial volume which consists of multiple tissues. In addition, the center intensity of the lesions is different from its border of lesions and similar to GM intensities. This phenomenon creates confusion between MS lesions and GM which leads to misclassification. It also happens with GM and CSF as they are on the border of the brain and which is normally misclassified as MS lesions. The damaging of brain tissue is not only limited to WM but can also affect GM. According to Trap et al \citep{trapp2008a}, 40-70\% of MS patients affected with lesions appear in GM. The segmentation of WM lesions is more difficult because of overlapping intensities with GM. In such scenario the T2-w images will not perform will due to its lack in contrast. However, FLAIR will perform better in GM lesions detection because of its high sensitivity.

\subsection{Types of MS Lesions}

Clinically isolated syndrome (CIS) is a term that refers to the patient’s first clinical visit with features that give indications of either a single lesion or multiple lesions and its symptoms remain for 24 hours. It’s not guaranteed that everyone who experiences CIS will develop MS. According to the National Multiple Sclerosis Society \citep{national_mult_sc_society}, there are different ways in which this disease acts:

\subsubsection{ Relapsing Remitting MS (RRMS)}
This stage is temporarily called relapses or exacerbation when its new symptoms appear. RRMS is the most common form of MS and around 85\% of people are initially affected by this type of MS.

\subsubsection{Secondary Progressive MS (SPMS)}
Those affected by RRMS are shifted to SPMS and its symptoms are more regular and worse than RRMS. The shift usually starts from 15 to 20 years from when a person is diagnosed with MS for the first time and about 80\% of people get SPMS.

\subsubsection{Primary Progressive MS (PPMS)}
This type is not as common as other types as only about 10\% of patients are affected by this type. It is characterized gradually with no relapses so it is hard to predict. PPMS affects the nerves in the spinal cord leading to problems in walking and balancing.

\subsubsection{Progressive Relapsing MS (PRMS)}

It is the rarest form of MS, about 5\% of MS patients have this type. It is characterized very slowly and worse than other types. The patient has acute relapses but no remissions.

\subsection{MS Lesions Localization}

The lesions are localized with different high or low intensities and the emergence of the necrotic area on edema. Periventricular lesions attached to both ventricular walls are normally larger in size, having a diameter of 5 - 10mm. Because of its large size, the detection of such lesions is easy, and it may have different shapes with high contrast on edema. FLAIR provides high sensitivity to detect such lesions because it can differentiate the signal from CSF \citep{casini2013a}\citep{s2008a}. Juxtacortical lesions are small with a spherical or elliptical shape and have low contrast. Due to these characteristics, the detection of such lesions is difficult because usually, the lesion size is smaller than the slice thickness \citep{garc2013a}. Cortical lesions consist of juxtacortical lesions located at the cortex which are adjacent to GM. They are easy to detect because of their contrast with GM. Necrotic lesions are in the center necrotic area with similar signals as GM, that’s why the necrotic area is mostly ignored for periventricular lesions.

\begin{figure*}
    \centering
    \includegraphics[width=1.0\textwidth]{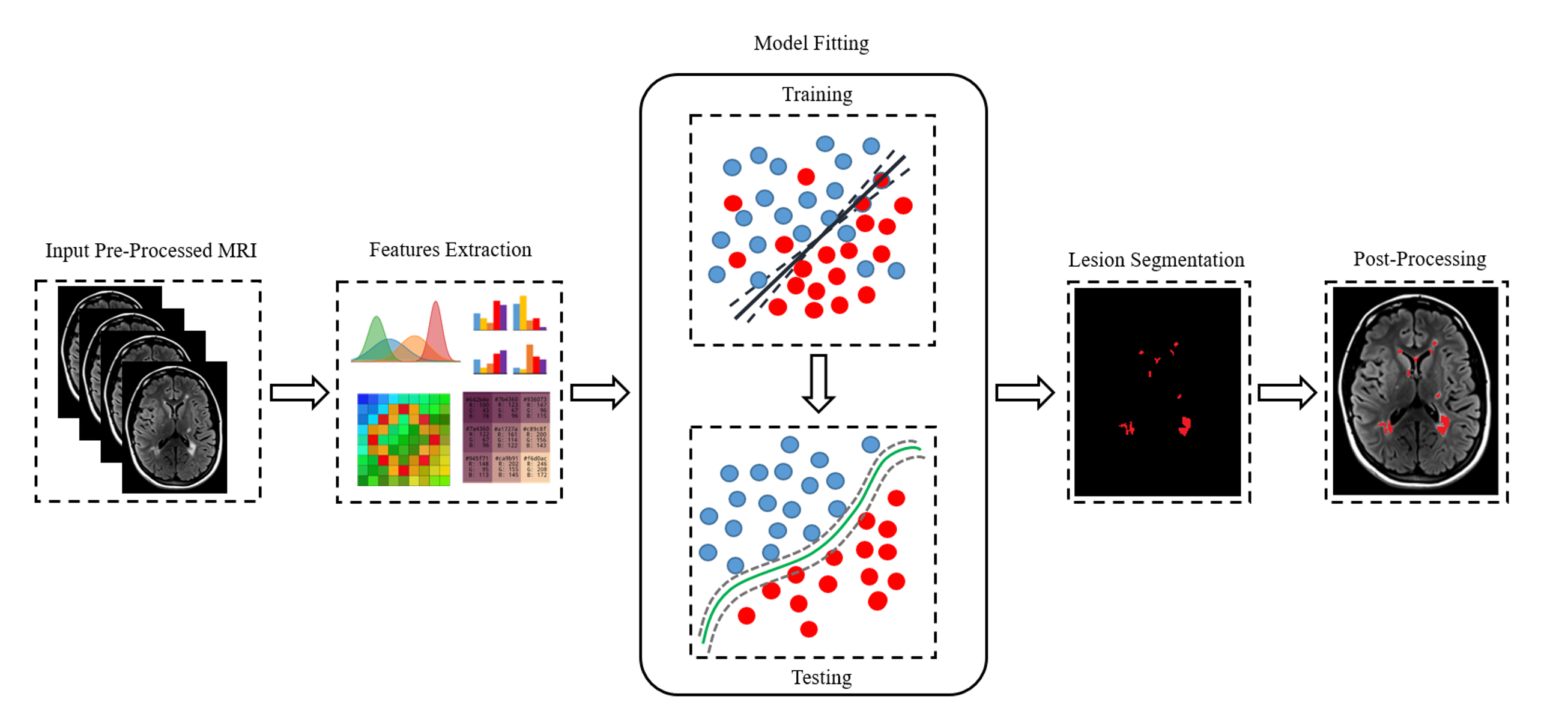}
    \caption{CAD system for MS lesions segmentation}
    \label{fig:cad}
\end{figure*}

\section{Generic Steps For MS CAD Systems}
\label{sec:3}
Different  CAD systems are proposed in the literature. They all share a generic sequence of steps. Figure \ref{fig:cad} illustrates the main steps in any CAD system which are discussed as follows:

\subsection{Image Acquisition}
This first step requires images which can be T1-w, T2-w, FLAIR and PD-w images which are already discussed in Section \ref{sec_sub:imaging}. Each image has its own characteristics such as the accurate classification of tissues. For example, GM and MS lesion have the same intensity in the T1 image thus classification will be a challenging task. Therefore, for that reason, some of the proposed algorithms focused on FLAIR images, in which MS lesions are more visible.

\subsection{Pre-Processing}

In the pre-processing step, the images are normalized according to the requirements. To increase the performance, the noise and inhomogeneity can be removed from the image which was generated during electric fields and scene capturing \citep{sled1998a}. Some other artifacts can also affect the segmentation process which includes blurred edges, motion, and overlapping intensities. At the first step, those artifacts, such as noise, will be removed. In the second step, those tissues, which are not related to the brain like the skull, will be removed. It is one of the necessary preprocessing steps for MS lesion segmentation. There are some reviews \citep{hou2006a}\citep{vovk2007a} on pre-processing which show that the results of segmentation can be affected if these artifacts are not treated well. However, there is a trade-off in pre-processing steps and algorithm complexity.

\subsection{Feature Extraction}
The focus of these techniques is the feature extraction from an image. Different features can be extracted by scanning an image column-wise, row-wise, or in a zig-zag manner in which scanning can be more productive. Different features can be combined to increase the performance such as combining the intensities of different images or different features of the same images. Principle component analysis can be used for feature reduction and feature dimensionality which decreases the computation complexity and increase the algorithm accuracy.

\subsection{Normalization}

Normalization refers to normalizing those features which can increase the performance. This step includes the removal of outliers and to have a mean of zero with a variance of 1.

\subsection{Classification}

The normalized features are fed to train the classifier. Unseen data can be used to check the performance of the classifier. Different classification algorithms are used like support vector machine, artificial neural network and fuzzy logic.

\subsection{Post-Processing}

Post-processing is used to improve the accuracy and reduce the true negative and false negative. These methods can be manual or automatic and are more reliable in some circumstances.  Contrast enhancement, spatial enhancement, histogram equalization and morphological operators are the main steps of post-processing.

\section{MS Lesion Datasets}
\label{sec:4}


Our literature review revealed several datasets that are publicly available and are popularly used in research.

\begin{table*}[]
\centering
\begin{tabular}[c]{|c|c|c|c|}
\hline
Institute                     & Scanner               & No. of Training Scans & No. of Test Scans \\ \hline
UMC Utrecht                   & 3 T Philips Achieva   & 20                    & 30                \\ \hline
NUHS Singapore                & 3 T Siemens TrioTim   & 20                    & 30                \\ \hline
\multirow{3}{*}{VU Amsterdam} & 3 T GE Signa HDxt     & 20                    & 30                \\ \cline{2-4} 
                              & 3 T Philips Ingenuity & 0                     & 10                \\ \cline{2-4} 
                              & 1.5 T GE Signa HDxt   & 0                     & 10                \\ \hline
\end{tabular}
\captionsetup{justification=centering}
\caption{WMH Segmentation challenge dataset}
\label{tab:table1}
\end{table*}

\begin{table*}[]
\centering
\begin{tabular}{|l|l|l|l|l|}
\hline
Dataset &
  No. of Patients (M/F) &
  Time-Points Mean (SD) &
  Age Mean (SD) &
  Follow-up Mean (SD) \\ \hline
Training            & 5(1/4)   & 4( $\pm$ 0.55)   & 43.5( $\pm$ 10.3 ) & 1.0( $\pm$ 0.13) \\ \hline
RR                  & 1(1/3)   & 4.4( $\pm$ 0.50) & 43.5( $\pm$ 10.3 ) & 1.0( $\pm$ 0.14) \\ \hline
\multirow{3}{*}{PP} & 1(0/1)   & 4.0        & 57.9         & 1.0( $\pm$ 0.04) \\ \cline{2-5} 
                    & 14(3/11) & 4.4( $\pm$ 0.63) & 39.3( $\pm$ 8.9)   & 1.0( $\pm$ 0.23) \\ \cline{2-5} 
                    & 12(3/9)  & 4.4( $\pm$ 0.67) & 39.5( $\pm$ 9.6 )  & 1.0( $\pm$ 0.25) \\ \hline
PP                  & 1(0/1)   & 4.0        & 39.0         & 1.0( $\pm$ 0.04) \\ \hline
PP                  & 1(0/1)   & 4.0        & 41.7         & 1.0( $\pm$ 0.05) \\ \hline
\end{tabular}
\caption{Longitudinal MS lesion dataset}
\label{tab:table2}
\end{table*}

\subsection{MICCAI}

Medical Image Computing and Computer-Assisted Intervention (MICCAI) (2008) MS Grand Challenge \citep{styner2008a} organizes various competitions, including White Matter Hyperintensity (WMH) Segmentation Challenge \citep{wmh_seg_challenge}. The data provided are collected from three different hospitals including Universitair Medisch Centrum Utrecht (UMC), The National University Health System (NUHS), Singapore and Vrije Universiteit (VU), Amsterdam. Table \ref{tab:table1} shows the information from different hospitals, scanner machines and the number of samples from the WMH Segmentation Challenge dataset.

\subsection{BRAINWEB}

BrainWeb \citep{cocosco1997a} dataset is a simulated dataset used to evaluate segmentation methods. The dataset uses synthetic images that make segmentation easy and these images have a lack of contrast such as FLAIR. These datasets are used to generate images based on user requirements and can be tuned. These images are also able to create various contrast images and can be used for evaluation. Recently one study \citep{hagiwara2017a} discussed the conventional and synthetic MRI to detect the MS plaques.

\subsection{LMSLSC}

The Image Analysis and Communications Lab (IACL) organized the Longitudinal MS Lesion Segmentation Challenge to evaluate different segmentation with Longitudinal Multiple Sclerosis Lesion Segmentation Challenge (LMSLSC) dataset \citep{carass2017a}. The data was acquired from real MS patients at a different time. Ground truth by two experts has been used for evaluation. Table \ref{tab:table2} shows the dataset information. The dataset is divided into two parts training and testing. Which consist of different MS types such that RRMS, PPMS, and SPMS. The table header shows the total number of patients while M/F shows the male/female ratio. Time-points and Age of patients are shown with their mean and standard deviation (SD). Follow up time in a year for MRI-scans are shown with their mean and SD. The dataset acquires via Tesla 3.0 MRI scanner. The images included in this dataset are T1-w, T2-w, PD-w, and FLAIR with a pre-processed image for each scan modalities.

\subsection{ISBI}

International Symposium on Biomedical Imaging Challenge (ISBI) 2015 \citep{int_symp_bio_img} provides a dataset for longitudinal MS segmentation. The training data consist of longitudinal images from five patients while the test data consist of two datasets from five and ten patients. Each dataset consists of T1-w, T2-w and FLAIR MR Images which are acquired using 3T MR scanner.

\begin{table*}[]
\centering
\def\arraystretch{3}
\begin{tabular}{|c|c|c|}
\hline
Name                           & Metric                                   & Definition \\ \hline
\multirow{8}{*}{Hard Measures} & Accuracy                                 & $\frac{|TN|+|TP|}{|TN|+|TP|+|FP|+|FN|}$ \\ \cline{2-3} 
                               & Sensitivity (True   Positive Rate (TPR)) & $\frac{|TP|}{|TP|+|FN|}$           \\ \cline{2-3} 
                               & Specificity (True   Negative Rate (TNR)) & $\frac{|TN|}{|TN|+|FP|}$            \\ \cline{2-3} 
                               & Positive Predictive   Value (PPV)        & $\frac{|TP|}{|TP|+|FP|}$            \\ \cline{2-3} 
                               & Negative Predictive   Value (NPV)        & $\frac{|TN|}{|TN|+|FN|}$           \\ \cline{2-3} 
                               & False Negative Rate                      & $\frac{|FN|}{|FN|+|TP|}$           \\ \cline{2-3} 
                               & False Positive Rate                      & $\frac{|FP|}{|FP|+|TN|}$            \\ \cline{2-3} 
                               & Dice similarity   coefficient (DSC) rate & $\frac{2 \times |TP|}{2 \times |TP|+|FP|+|FN|}$     \\ \hline
\multirow{3}{*}{Probabilistic} & Probabilistic   similarity index         & $\frac{2 \times \sum P_{x,gs=1}}{\sum 1_{x,gs=1}+\sum P_x}$           \\ \cline{2-3} 
                               & Probabilistic overlap   fraction         & $\frac{\sum P_{x,gs=1}}{\sum 1_{x,gs=1}}$           \\ \cline{2-3} 
                               & Probabilistic extra   fraction           & $\frac{\sum P_{x,gs=0}}{\sum 1_{x,gs=1}}$           \\ \hline
\end{tabular}
\caption{Evaluation Metrics}
\label{tab:table3}
\end{table*}

\section{CAD Evaluation Measures}
\label{sec:5}
CAD systems have been evaluated with different measures \citep{llad2012a} \citep{petrick2013evaluation}\citep{chabi2012evaluation}. However, all these measures compare the results of segmentation methods with ground truth, annotated by deep observers and experts. The ground truth of more than a single expert should be used to avoid the inter-observer and intra-observer variability because this contributes to the more persistent ground truth. Some approaches like STAPLE \citep{warfield2004simultaneous} fused these annotations from different experts. The segmentation and ground truth can be compared voxel by voxel, or it can be compared with a whole segmented lesion. In both cases, voxels are classified as True Positive (TP), True Negative (TN), False Positive (FP) and False Negative (FN). In medical applications, the objective is to reduce the number of FPs and FNs and maximize the number of TPs and TNs. However, there is a trade-off between these two values. By the increase of TPs, the FPs also increase and decreasing the TPs will decrease the FPs. However, an increase in the FPs also reduces the confidence level which can be calculated via statistical tools. DSC rate and the Jaccard Similarity Index (SI) are also used in these systems. These metrics show the overlaps value with the ground truth, the closer the value to 1, the more accurate the segmentation will be while the closer the value is to zero the worse the segmentation. Table \ref{tab:table3} shows all the evaluation metrics with mathematical definition.

\section{Methodology}
\label{sec:6}
This section discusses our research methodology for conducting the review.  The research questions, search inclusion criteria, search strategy, and extraction of studies and findings are as follows.

\subsection{Research Questions}

This study considers three research questions:
\begin{itemize}
    \item RQ1: What techniques have been used in computer-aided diagnosis and classification of MS lesions? What are their advantages and limitations?
    \item RQ2: What are the different techniques used for pre-processing and features extraction?
    \item RQ3: What are the Machine and deep learning approach's for MS classification and segmentations?
    \item RQ4:  What are the challenges/gaps to be addressed in future research?
\end{itemize}

\subsection{Inclusion Criteria}

Studies included in this paper should fulfill the following criteria:

\begin{itemize}
    \item Present the techniques which segment and/or classify MS lesions.
    \item Methods using computerized system and machine learning techniques.
    \item Evaluate some quantitative measure of medical images.
    \item Present adequate results with evaluation metrics.
    \item The papers should be published in peer-reviewed journals, conference proceedings and a book chapter.
    \item The paper should evaluate brain MR Images to detect MS lesions.
\end{itemize}

\subsection{Search Strategy}

To find state-of-the-art CAD systems that classify MS lesion, various search engines and databases were used as follows: Science Direct, IEEE Xplore, Google Scholar, Scopus, and ACM digital library. Those studies which were published from 2001 -- 2020 (excluding which reviewed by previous journals) were taken into consideration with more focus on 2010 -- 2020. The vocabulary used for different search keywords include: ``Multiple sclerosis segmentation", ``White matter lesions", ``MS lesion detection", and ``Computer-Aided Diagnosis system for MS lesion".

\subsection{Extraction of Studies and Findings}

Table \ref{tab:table-img-seq} illustrates the extracted data from various studies which shows: Authors name, Publication year, image sequence, and what type of lesions were detected. Table \ref{tab:summary} shows more in-depth extracted information which includes, algorithm and methods, approaches, the datasets and the evaluation metrics to evaluate its methodology.

\section{MS Lesion Segmentation and Detection Techniques}
\label{sec:7}
\begin{figure}
    \centering
    \includegraphics[width=3in]{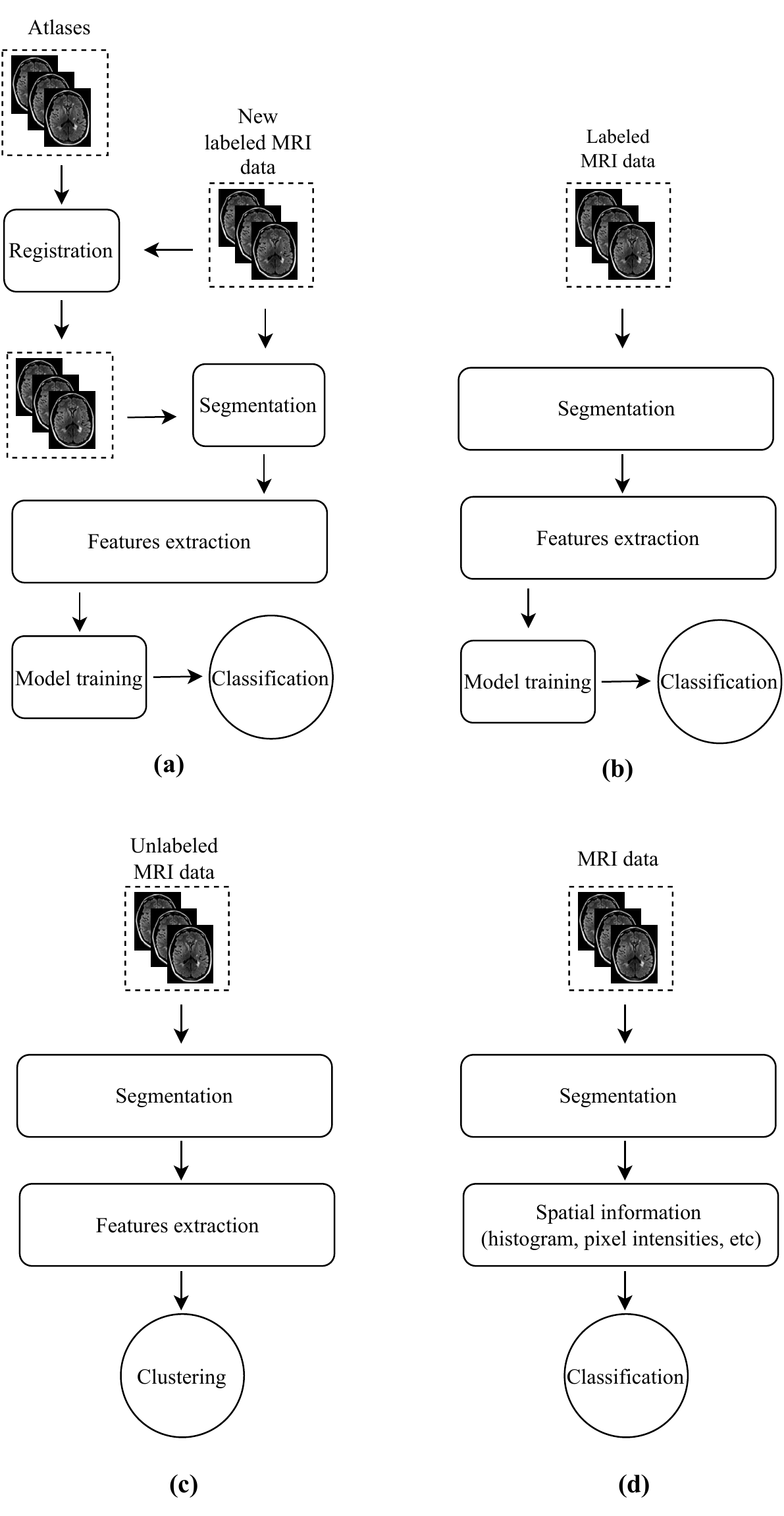}
    \caption{Flow charts of (a) Supervised learning with atlases (b) Supervised learning (c) Unsupervised learning and (d) Data-driven methods }
    \label{fig:cat}
\end{figure}

The techniques used in MS lesion detection follow different approaches. The criteria to classify these approaches, according to the literature, is based on the methodology. In this study, we categorize MS lesion segmentation techniques into six categories: Data-driven methods, Statistical methods, Supervised methods, Unsupervised methods, Deep learning methods, and  Fuzzy methods. It is noteworthy here that we distinguish between the Data-driven, Supervised, and Unsupervised methods, Figure \ref{fig:cat} illustrates the difference. 
Data-driven methods describe those methods which use spatial information, voxels neighborhood information, intensity-based and histogram-based information. 
Supervised learning methods include those methods which may need some prior information (atlases) and labeled data to train the learning algorithms. Figure \ref{fig:cat} highlights the differences among these methods with and without atlases.
Unsupervised methods include those methods which do not need prior information nor label data, for example, clustering techniques. 
Statistical methods include those approaches which use probabilistic models, probabilistic atlas, expectation-maximization and other statistical approaches. 
Deep learning methods, which come as an extension to machine learning algorithms, include those methods which use deep neural network architecture to train the algorithm with a large amount of data to improve the performance. 
The fuzzy methods uses fuzzy inference rules for classification.

\subsection{Data-driven Methods}
The automatic segmentation of MS lesion methods was proposed in \citep{khayati2008novel}. Three different MR images were used: T1, T2, and FLAIR. In T1 the lesion which has hypo-intensities are called black holes, while in T2 they are hyper-intensities. There are three different types of lesion stages: acute or enhanced in T1, chronic in T1, and other in T2 images. The acute lesions are further divided into two: the new lesion with a higher activity called early acute while minimal activity is called recent acute. So, in this work, they automatically differentiate the different stages of MS lesions based on the voxels’ intensities. First, lesions are extracted from normal tissues. In the second step, a generated mask from a lesion class is applied to FLAIR to extract lesions. The third step is to map voxel activity. In the fourth step, the voxel results are mapped to classify the chronic and acute lesions by optimal thresholding, staging, and automatic staging. To evaluate the results, it was compared with manual segmentation.

Liu et al \citep{liu2009a} extracted texture features of gray level pixels to diagnose MS lesions. Four different texture features were used including contrast, angular second moment, mean and entropy. T2-weighted MR images were used from 26 patients and 26 healthy subjects. The texture features are extracted from the region of interest to achieve better results. The ANOVA was applied to evaluate the results.

Roura et al \citep{roura2015a} extended the work in \citep{cabezas2014a} and improved biased segmentation, tissue segmentation steps and lesion segmentation. Through this improvement, the number of TPs remained the same, but the FPs detection rate was reduced.  In the proposed method, the preprocessing step involves skull stripping and noise removal The main focus of the work is to segment the lesion based on their intensities and for that, intensity normalization is a crucial step for which N3 intensity correction is used in \citep{popescu2012a}. They used re-slice and spatial co-registration methods via SMP8 tool. Different methods were used in thresholding to remove FPs and make them restricted while on the other side it increased the TPs.

Villalon-Reina et al \citep{villalon_reina2013a}  developed a bio-inspired artificial visual model to detect MS lesions by using FLAIR images. In pre-processing steps, Brain Extraction Tool (BET) \citep{BET} is used to remove the skull and extract the brain tissues and the ITK-SNAP tool \citep{ITK_SNAP} is used to smooth the images. The visual model consists of various filters called model-cells. These filters consist of various stages including, lateral geniculate nucleus, V1 sample, complex cells, V2 and V4 area. The results were evaluated with sensitivity, specificity, and dice coefficient performance measures.

Hill et al \citep{hill2014a} proposed Improved Jump Method (IJM) to segment MS lesions by using pseudo-color images.  Preprocessing steps consist of rescaling brain data, removing the skull and smoothing images. First, the brain image was divided into four clusters by using the quality of a specific cluster in the optimization step. IJM estimates the distortion rate curve and spectral clustering approach \citep{hadjighasem2016a} are used to address the nonlinearity. Post-processing is used to segment the CSF and MS lesions from the IJM image segmentation, and this segmentation is done via the threshold of the pseudo-red components.

Schmidt et al \citep{schmidt2012a} automatically detected a T2-hyperintense WM lesion through T1-w and FLAIR images. Inhomogeneity was corrected, and edges were smoothened in the pre-processing step. GM, WM and CSF tissues classes were extracted by using partial volume estimation and thresholding. Intensity distribution was calculated from these three tissues to detect FLAIR outliers. The WM belief maps were used as an initial region for growing seeds for lesions, which evaluate their neighbor voxels and classify them as lesions or normal tissues iteratively. The approach was evaluated via the correlation between automatic and manual segmentation with performance measures of sensitivity, specificity, and accuracy. 

Zhong et al \citep{zhong2014a} proposed a pipeline to extract the WMH lesions, calculate their volume, count, and categorization based on their locations. The proposed method consists of various steps; skull stripping and bias correction of inhomogeneity in the pre-processing step, ventricle and CSF extraction using connected pixels, WMH and small lesion segmentation using high-intensity thresholding and the last step including lesion volume extraction via region growing algorithm. The ventricle information was used to categorize the lesions. Lesions having edges less than 3.0 mm are classified as periventricular hyperintensities while larger edges are classified as Deep WMH.

Ganiler et al \citep{ganiler2014a} proposed a subtraction pipeline to detect new lesions in longitudinal scans. Initially, the skull was extracted using BET. PD-w images were used for masking, and bias correction was applied to these masked images at all time points. The intensities of scans were normalized via histogram matching approaches and high-intensity values were normalized to obtain better results. The registration was applied to the normalized images. The registration was done with mutual information metric \citep{mattes2001a} which is used as the cost function. Gradient descent optimization is used to minimize the cost function.

Zeng et al \citep{zeng2014a} proposed a technique to segment lesion and reduce the FPs. Brain tissues were extracted using the BET tool. These tissues were used as a binary mask. To enhance the density of MS, lesion T2 and FLAIR were fused. Gaussian Markov model is used to segment tissues into CSF, GM, WM and background. The image enhancement function used to enhance MR images, and mutual information (threshold) was used to get successive iteration and stopping criteria. Alpha matting \citep{levin2008a} was used to refine the results of lesion segmentation.

Wiggermann et al \citep{wiggermann2016flair2} combined FLAIR and T2 to obtain FLAIR2 images to detect WM lesions with a better detection rate. The contrast to noise ratio was improved between lesions, WM, and GM as compared to conventional MR images. 3D FLAIR2 was also used to better visualize callosal and infratentorial lesions.

Storelli et al \citep{storelli2016a} used semi-automatic segmentation where the lesion was manually segmented for training via experts. The images were divided into groups, a low lesion load was selected from each group and a threshold function was defined, which can refine these groups for training. Regions growing algorithm was used to segment lesions. Different line fitting models were used and the one with maximum DSC was selected.

Revenaz et al \citep{revenaz2016a} used various techniques hierarchically to segment various tissues. The steps involved were, preprocessing, co-registration 2D and 3D in FLAIR, segmentation of lesions and normal tissues and usage of T1 images to get a mask for FLAIR lesions.

Mechrez et al \citep{mechrez-a} proposed a method based on the similarity of multi-channel patches. A database of patches is used to compare each patch, which was labeled as a training map. An iterative patch-based refinement process is used to segment the patches until the detection of lesions is consistent. The method is evaluated on MICCAI 2008 dataset \citep{styner2008a} where the training section of the dataset was used for training patches and the testing section was used for the testing patches.

Cabezas et al \citep{cabezas2016a} used to detect the new lesions in follow-up MRI images using deformation fields. In the pre-processing step, brain tissues are extracted using BET, the brain mask is extracted, the N4 algorithm is used to correct the biases and lastly, the histogram matching approach is used to normalize the intensity of MRI images. The affine transformation is used for image registration and a demon’s algorithm is used to obtain the deformation field. Image subtraction and thresholding are applied to the lesion mask and refined via the deformation field.

Storelli et al \citep{storelli2016b} presented a semi-automatic lesion segmentation method based on the manual identification of lesions in dual-echo images. The region growing algorithm was applied to the manually identified lesion to get the thresholding which can push or stop the algorithm from growing. This process was repeated to get the optimal threshold value. This threshold value was applied to the test images to extract MS lesions.

Roy et al \citep{roy2017a} used thresholding and binarization to detect lesions. Contours of the brain were used to connect normal tissues to a near boundary. Then all the lesions were removed except maximum connected area lesions. The binarization and thresholding were done based on the standard deviation and entropy after image enhancement. The background image was subtracted from the binarized image to extract the MS lesions.

Egger et al \citep{egger2017a} compared two automated algorithms with the manual segmentation on 50 MR images acquired from The Brain and Mind Center, University of Sydney, Australia. The two automated algorithms were the lesion prediction algorithm , available in the SPM12 tool, and the Lesion Growth Algorithm (LGA), available in the SPM8 tool.  Manual segmentation was performed by three independent expert raters. The results were evaluated via DC and inter-rater correlation coefficient.

Chen et al \citep{chen2017a} used Diffusion Tensor Images (DTI) to detect occult lesions in MS. The parameter they estimated was anisotropy, meaning diffusivity, and did the comparative analysis with 10 MS patients and 10 normal people. The analysis shows that, in acute lesions, the anisotropy values were decreased while the mean diffusivity values were increased. MS lesions reduce the color value signals and also appear larger in DTI as compared to Diffusion-weighted Images.

Nafisi et al \citep{nafisi-moghadam-a} compared Diffuse Weighted Images (DWI) and FLAIR to detect MS plaques/lesions. The study showed that the MS lesion diameter is smaller in DWI as compared to FLAIR.  Also, more lesions were detected in FLAIR as compared to DWI which shows that FLAIR is more efficient to detect MS lesions.

Andermatt et al \citep{andermatt2017a} proposed a semi-automatic framework to track the evolution of an individual MS lesion and also estimate proportion enhancing lesions. The framework was evaluated via 50 patients’ data which shows that the similarity between rater and framework is greater than 98.
Zeng et al \citep{zeng2013a} proposed a method based on 3D volume lesion segmentation. This approach consists of three steps, in the pre-processing step, the mutual information approach is used for the registration of T2-w and FLAIR images using skull removing and image enhancement. In the second step, to segment, the lesion, correctly, removes false positives via symmetry plane estimation. In the third step, the alpha matting method was used to increase segmentation accuracy.

Eichinger et al \citep{eichinger2017a} detected and visualized new MS lesions using Double Inversion Recovery (DIR) image sequence which might help boost clinical work and increase the sensitivity in DIR images. The study followed three subtraction maps including the standard visualization comparison, FLAIR subtraction maps and DIR subtractions maps. The DIR subtraction maps lead to high accuracy and detect a significantly large number of new lesions than the other two methods.

Schmidt et al \citep{schmidt-a} used image registration and fusion with post-processing contrast adjustment to detect new MS lesions. The results were evaluated by four experts and showed significant sensitivity of 100.

Stamile et al \citep{stamile2017a} used Genetic Algorithm (GA) to modify the histogram distribution to detect longitudinal changes in WM fiber of patients with MS. 

An automatic technique proposed by Rodrigo et al \citep{rodrigo2015a} classified the lesions based on image intensities. This method consists of four steps:
\begin{itemize}
    \item histogram specification, which is used to normalize the intensities of the lesion.
    \item brain segmentation, which is done to remove extra-cerebral structures such as the skull and scalp.
    \item the volume of the dataset is co-registered with gray matters which helps the segmentation procedure (Three different probabilistic classes are used to segment each voxel as WM, GM, and CSF).
    \item thresholding is applied to get the final results, but before thresholding,  image enhancement techniques were used with a high pass filter and gamma correction.
\end{itemize}

Authors in \citep{zhao2018a} used a three-phase level and bias field estimation to segment tissues and MS lesions. Initially, the experiments used T1-w and FLAIR images to segment lesions and GM, WM and CSF. In the second step, the phase level approach is used to improve segmentation accuracy. 

Ghribi et al \citep{ghribi2018a} extracted three types of features Volumetric Gray Level Co-occurrence Matrix (GLCM), Volumetric Gray-level Run Length Encoding Matrix (GLRLM) and Volumetric Shape Matrix. The GLCM consist of texture and second-order statistical features. Genetic algorithm-based Support Vector Machine (SVM) was used to extract the optimized features and do the segmentation accurately.

González-Villàa et al \citep{gonzalez2019brain} used a semi-automatic intensity-based multi-atlas method to segment MS lesions. Two strategies were used in the proposed work are joint label fusion and Non-Local-Spatial STAPLE (NLSS). 

Schmidt et al \citep{schmidt2019a} use serial MRI to detect changes in WM lesion in MS. The intensity-based method was used with respect to two-time points. The intensity was analyzed between two-time stamps to segment lesions. The results show that voxel-wise DSC achieved 0.7 and lesion-wise rate of 0.8 with 0.2 false-discovery rates.

Pelizzar et al \citep{pelizzari2020semi} devised a semi-automatic method for detecting MS areas of increased susceptibility in white matter (WM) lesions. The method primarily defines an intensity threshold for the processed Quantitative Susceptibility Maps (QSM) and Susceptibility-Weighted Imaging (SWI) to mark the susceptibility areas. The dataset was collected from thirty-eight MS participants. 

\subsection{Statistical Methods}

This section consists of those methods which uses statistical models, estimation techniques and probabilistic techniques to classify MS lesions. 
karimaghaloo et al \citep{karimaghaloo2013a} proposed a method that takes in robust higher-order texture and local voxels level pattern into the model. This work is based on Conditional Random Field (CRF) with a modified adaptive step. The Adaptive Multi-level CRF (AMCRF) classifier uses the temporal and spatial information to detect MS lesions. The CRF develops binary labels and voxels with the same labels are grouped, each group is analyzed more to extract textural patterns, and RIFT features \citep{lazebnik2005a} are used because it’s invariant to intensities and rotation. To reduce the false lesion, AMCRF was used which also refine lesion boundaries.

Bilello et al \citep{bilello2013a} developed a CAD system that can track temporal changes of MS lesions in serial MRIs. The pre-processing step involves skull removing and inhomogeneity correction. Candidate lesions were extracted from MRIs. Linear regression was used to extract the range of intensities of a particular scan. Based on these intensities, the CAD system threshold dynamically removes false positives.

Fully automatic Bayesian work was designed for MS classification by Harmouche et al \citep{harmouche2006a}. The lesions were classified by using posterior probability with entropy values. The regional specific multivariate likelihood was used to model the variability of intensities in various multi-model MRIs. Markov random fields were applied to remove noise and use a neighbor voxel to smooth the segmented regions. The method shows acceptable results to classify the cerebellum which is nearly comparable to ground truth.

Bricq et al \citep{bricq2008a} used the Hidden Markov Chain model for neighborhood voxel information and extracted the outliers by using Trimmed Likelihood Estimation (TLE) \citep{royall1986a}. This method uses 3D multi-modal MR images to detect MS lesions. The approach was evaluated in four different cases with the mean SI of 0.582.

Parastwa et al \citep{m2008a} proposed a method that detects lesions as outliers through an atlas without requiring any training data. The Region-based classification was done instead of voxel-based to avoid false lesions and detect real lesions. Probability Density Function (PDF) for intensity was calculated for healthy tissues and the Minimum Covariance Determinant (MCD) was used for outlier discrimination.

Spatio-temporal Robust Expectation-Maximization (STREM) was proposed by Ait-Ali et al \citep{a2005a} to detect MS lesions robustly from 3D MR images data. Maximum Likelihood Estimation (MLE) method was replaced with Trimmed Likelihood Estimator (TLE) and its parameters were used iteratively to detect MS lesions. Prior information and Mahalanobis distance were used to refine the segmented data.

Garcia-Lorenzo et al \citep{garc2008a} also used STREM for MS lesion diagnosis. The pre-processing step involves inhomogeneity correction, de-noising, and skull stripping. The MS lesion segmentation process involves a robust estimation of Normal Appearing Brain Tissues (NABT) followed by outlier detection. A Modified Multivariate Gaussian mixture model was used to calculate NABT parameters. The modified version was based on TLE. The proposed method was evaluated on two datasets from two different hospitals with a mean specificity of 0.995 and a sensitivity of 0.256.

The Constrained Gaussian Mixture model (CGMM) was proposed by Freifel et al \citep{freifeld2009a}. Multiple Gaussian distributions are used for each brain image including T1, T2, and PD. EM is used to estimate the parameters of a model. Detected MS lesions are considered as outliers. To refine the lesion boundaries, the probability-based curve evolution approach is used. The model does not require atlas for its initial learning. CGMM curve evolution model was evaluated on simulated images and also with a real dataset with highly noisy data.

Gao et al \citep{gao2013a} proposed an energy minimization based technique to detect lesions by using T1-w, T2-w and FLAIR images. The multi-channel MR images are used for this model, and energy minimization is used to get the tissues’ optimal membership function. In the preprocessing step, intensity normalization bias correction is done. The method is robust to intensity inhomogeneity and noise, so the segmentation is not affected by image modality and lesion type.

Lie et al \citep{liu2009b} proposed a fully automatic MS lesion segmentation by using T1, T2, and FLAIR images. The EM approach is used to separate the background of different tissues such as CSF, WM, and GM. The lesions are classified as outliers, and the L2E measure is minimized to separate the lesions from normal tissues. False positive pruning is used to refine the accuracy. 

Sajja et al \citep{sajja2006a} proposed the method involving the combination of parametric and non-parametric approaches which reduce false MS lesion classification. Contextual information, used through hidden Markov random fields with expectation maximization (HMRF-EM), is used to minimize false negative classification. The algorithm detection rate is 80\% with low-lesion load and 93\% in high lesion load.

The Bayesian classifier was proposed by Khayati et al \citep{khayati2008a} which automatically segments MS lesions. Adoptive Mixture Method (AMM) is used to estimate the probability density function for each class. The prior probabilities are calculated via Markov Random Field (MRF). The parameters of the classes like the mean and the variance are updated without any training. By thresholding, combining prior probability and class condition probability, the fully automatic segmentation is done. To evaluate these results, using similarity criteria, overlap fraction and extra fraction are calculated showing better results than previous work.

Shiee et al \citep{shiee2010a} used topological and statistical atlas-based approaches that classify WM and GM lesions. The method used manual align atlas for the initial segmentation step and computed weights and membership. The steps were repeated to refine the results. Real-time images were used including T1, T2 and FLAIR to evaluate the algorithm.

Harmoche et al \citep{harmouche2015a} proposed a probabilistic method to detect the intensity and spatial information based on MS lesions in different MR images. It segments different regions of brain images based on similar intensities. These intensities are modeled into different tissue classes via multivariate Gaussian like the central brain, frontal lobe, occipital lobe, temporal lobe, parietal lobe, and posterior fossa, which are different regions of the brain. MRF and local neighboring voxels are considered to make it more efficient for classifying tissue classes.

Spies et al \citep{spies2013a} detected hypointense lesions by using T1 images. Hypointense WM lesion intensities are similar to GM in T1 images. So, the first step of this approach involves separating GM, WM, and CSF. The GM lesions probability map is tested on each voxel against healthy GM maps. GM with high density in Region of Interest (ROI) for WM is clustered together and marked as a lesion. Each lesion contours are transferred to co-registered FLAIR and T2 images of the same patient to detect corresponding T2 hypointense lesions. The algorithm used simulated and real datasets and DC. Accuracy and FP performance measures were used for the evaluation.

Karpate et al \citep{karpate2014a} proposed a method to segment longitudinal MS lesion using T1-w, T2-w and FLAIR images. Intensity standardization is used to minimize the differences in inter-scan intensities. To obtain the intensity normalization maps, Gaussian Mixture Model (GMM) estimation is used, and to estimate loss function, MLE is used. Gong et al \citep{gong2015a} used robust energy minimization based approach to detect lesions by using FLAIR images.

Karimaghaloo et al \citep{karimaghaloo2015a} used texture features in a hierarchical with CRF to classify MS lesions. They call their method Temporal Hierarchical Adaptive Texture CRF, THAT-CRF. The method consists of various steps, CRF based classifier is used to detect the candidate lesions in the very first step. CRF is used with a high order clique of three which can capture more complex information in images. Temporal and stationary high order features were extracted from patches that contain lesions. Both features were evaluated independently and also combined, and the robustness of LBP was explored using local intensity histogram and rotational invariant feature transform (RIFT). The CRF is used with voxel-wise features and high order texture features to remove false detected regions. Sensitivity and false detection rates are used for performance evaluation.

Tomas-Fernandez et al \citep{tomas-fernandez2015a} collected data from patients including intensities and segmented templates of brain tissues which are referred to as the Model of Population and Subjects (MOPS). This approach is unlike classification, but it compares abnormal tissue intensities with expected values in populations of healthy tissues in the same location. This population can be achieved via the combination of GMM and the local tissue intensity model. The model was evaluated on both synthetic and clinical datasets.

Elliott et al \citep{elliott2013a} proposed a probabilistic approach to detect new lesions by using follow-up scans. The method not only detects new lesion time points but also estimates the change due to artifacts or miss-registration. The approach consists of two main stages. First, the Bayesian model is used on each voxel by using follow up scans. This will yield a voxel-wise probabilistic classification. Second, the new lesions are added to a group of candidate lesions with a confidence value assigned via random forest classifier. The Bayesian classifier uses spatial and neighborhood information while the random forest refines the initial voxel classification by using new candidate lesions.

Gao et al \citep{gao2014a} proposed a novel approach to segment lesions in multi-channel MRIs. The energy minimization approach was used for MS lesion segmentation and the bias field was estimated from multichannel MRIs. To achieve regularized segmentation and to reduce noise, the non-local mean approach was used. To evaluate the algorithm pipeline DS, FPR and specificity performance measures are used.

Freire et al \citep{freire2016a} used the iterative segmentation method using a student t mixture model with probabilistic atlases. The brain tissue extraction, noise removing, bias field correction and image registration were done in the pre-processing step. The method used intensity-based segmentation via the student t model and refined the segmentation iteratively. The main steps of this approach are: 
\begin{itemize}
    \item in the pre-processing step, a binary mask was created via atlases.
    \item Defined the number of clusters for each iteration.
    \item Segmented the image using the student t model with the binary mask and defined clusters.
    \item Calculated the mean for each cluster.
    \item Used the highest mean intensity of the previous cluster as a binary mask for next iteration.
    \item Repeated step 3 until iteration reached to zero.
\end{itemize}

Galimzianova et al \citep{galimzianova2016a} used Stratified mixture modeling to segment WML. The method uses the distribution of intensities of normal-appearing brain structures to detect lesions as outliers. This approach is incorporated with three unsupervised previous methods  \citep{leemput2001a}, \citep{garc2009a}, \citep{garcia-lorenzo2011a}  which used GMM with Stratified mixture modeling, approving the results of accuracy and reducing FPs.

Karimaghaloo et al \citep{karimaghaloo2016a} used probabilistic multi-level AMCRF with high order cliques to segment and detect MS lesions. The method has two main steps. First, the voxel-based CRF model was used to extract the candidate lesions. Patches were extracted from those identified lesions to use the contextual information. To increase the robustness of the model, the Relevance Vector Machine (RVM) texture classifier was trained using the candidate lesions. The AMCRF reduced the number of FPs but also adaptively changed the boundaries of the lesion which can best fit the ground truth.

Strumia et al \citep{strumia2016a} used the Geometric brain model instead of Atlas to segment WMLs. They used the connectivity of GM and relative locations instead of the various locations of tissues. The model achieved the results without artifacts which can be introduced via Atlas.

Zangeneh and Yazdi \citep{zangeneh2016a} used constrained GMM and GA to automatically segment the MS in brain MRI. The pre-processing step consists of skull removing, noise and artifacts removing and the normalization of the image to preserve the initial image range. The GMM is used to classify the image pixels into four classes. Five various constraints are defined which also include the intensity of the MS pixels which will be higher than 0.85 in normalized images. The Genetic algorithm was used to solve the nonlinear constrained for the GMM.

Puonti and Leemput \citep{puonti2016a} developed a novel method that can be combined with existing segmentation methods to segment the whole brain in MRI images. The work is focused on the Boltzmann machine which gives richer spatial models. The proposed method was evaluated via MICCAI 2008 \citep{styner2008a}.

Ziga et al \citep{lesjak2016a} used longitudinal MR images of 20 patients to achieve a quantitative and systematic evaluation of supervised and unsupervised intensity-based methods. These methods were used to detect changes in intensities in follow-up MR images. The authors also focused on pre- and post-processing steps to evaluate the metrics.

Jain et al \citep{jain2016a} segmented two times MS lesions via using the expectation maximization approach. This pipeline consists of three main steps: In the first step, cross-section analysis of the lesion; WM, CSF, and GM were classified as outliers to normal brain tissues. In the second step, FLAIR difference images were extracted via subtraction of FLAIR images after intensity normalization, biased correction, and co-registration. In the final step, a joint EM algorithm is used to optimize lesion segmentation by using the results of the first and second steps.

Liu et al \citep{liu2016a} improved the Spatial Regression Analysis of Diffusion tensor imaging (SPREAD). Previously, SPREAD used re-sampling and spatial regression techniques to detect changes in brain images, which also affects the boundaries and sensitivity. By using a nonlinear anisotropic diffusion filtering approach, edges can be preserved after filtering which improves the sensitivity and accuracy in lesion detection.

Zhao et al \citep{zhao2017a} proposed an improved version of multiplicative intrinsic component optimization (MICO) multichannel, where before, it was only limited to single channel MR images. In the pre-processing step, skull stripping, and registration were done. The membership function was updated via the energy minimization method until the convergence criteria were met.

Valverde et al \citep{valverde2017a} used T1-w and FLAIR images to detect white matter lesions by classifying white matter as outliers. White matter outlier filing and rejection, morphological and probabilistic prior maps and the combination of intensities were used to segment WM lesion. The performance was evaluated via MS patient scans and MRBrainS2013 dataset \citep{mendrik2015mrbrains} consisting of 20 scans with white matter lesions.

The study \citep{dworkin2018a} developed a method to count lesions with the correlation of age and lesion size. The method found that the count was highly correlated up to 97\% with age. The study also found a large number of small size lesions have better clinical outcomes that a small number of large size lesions. The method used the combination of T1-w, T2-w, FlAIR and PD images for the probability estimation of 60 MS patients.

Ghribi et al \citep{ghribi2019a} used a GMM to process various MRI modularity to segment MS lesions. The methodology uses adapted-GMM using an atlas and used region growing algorithm. The intensity, texture features and covariance among pixels were also taken into consideration in this work.

Wang et al \citep{wang2019a} first focused on the classification of WM, GM and cerebrospinal fluid and then segment lesion in WM regions using an atlas. The multi-atlas and multi-channel (MAMC) approaches were adopted and evaluated using MICCAI 2018 dataset.

Pota et al \citep{pota2019a} extracted volumetric and relaxometric data from MRIs. Statistical and fuzzy logic was used to segment lesions. Two models were generated with one and multi-dimensions to evaluate robustness, interpretability and generalizability of the proposed method.

Freire and Ferrari \citep{freire2019a} used hyperintensity probability maps to highlight hyperintensities lesion and reduce the intensities of WM and GM regions. The study found that using FLAIR the intensities of lesions were 19\% and 25\% brighter than GM and WM regions while applying this approach the enhanced version showed 264\% and 444\% brighter. 

Wang et al \citep{wang2020a} adopted the sparse Bayesian model to segment WM lesions using T1-w and T2-w and FLAIR images. This worked two approaches, first the sparse Bayesian theorem with MRF and Gibbs random field are used for WM lesion segmentation. In the second approach, the probabilistic label fusion algorithm was used based on a weight-based voting strategy. The MICCAI 2008 dataset was used for benchmarking with TPR, PPV and DSC as evaluation metrics.

Dufresne et al \citep{dufresne2020joint} introduced an optimization framework to track MS evolution. The framework tries to alleviate images deformable registration caused primarily by brain atrophy by jointly estimates the lesion change along with deformable registration as an optimization problem. The framework was evaluated on synthetic and real datasets against its counterpart sequential methods. The joint method outperforms the sequential methods on Dice metric. It performs 0.7, 0.7 respectively on two lesion evolution cases on the synthetic dataset and 0.52 on the real one.

\subsection{Supervised Methods}
In supervised based methods, the algorithms learn from previous data and are trained through training data. These algorithms can be evaluated via test data which consists of unseen data examples. The more the training data, the better the algorithm will learn. However, due to noise in data, it may lead to over-fitting.

Roy et al \citep{roy2013a} used the combination of local and global neighborhood texture features for the segmentation of MS lesions.  Robust lesion contrast enhancement and intensity normalization filters were used to enhance the region of interest. A support vector machine was used to classify the lesion regions. The post-processing step involved morphological filtering in achieving high accuracy.

Deshpande et al \citep{deshpande2015a} proposed a sparse representation and adaptive dictionary learning techniques to segment the MS lesion in MR images. Due to the size of the dictionary, a single dictionary represents all classes so that it might lead to the worst classification. Four different dictionaries are created, including a two-class dictionary with the same size or a two-class dictionary with different sizes. There is a four-class dictionary with the same size and a four-class dictionary with a different class (size). The two-class dictionary represents only lesion and non-lesion classes, while the four-class dictionary represents WM, GM, CSF, and lesion classes.  Experiments showed that the two-class with the same dictionary size has a high sensitivity rate and a positive predictive value as compared to the other three dictionaries. 

G´omez et al \citep{g2015a} segmented MR lesions in reconstructed multi-contrast MRIs. The joint reconstruction approach was used under different noise levels to show its robustness to noise. The images were corrupted with Rician noise and reconstructed using a joint reconstruction approach with two different parameters: diffusion and kurtosis model. Five different noise levels were used including, 0\%, 2\%, 4\%, 6\%, 8\% and 10\%. The lesion was detected and segmented in all these noisy MRIs and random forest algorithms were used for classification.

Texture features used by Zhang et al \citep{zhang2007a} from the GLCM were computed to classify normal tissues and MS lesions. The four GLCM texture features include contrast, the sum of a square, variance, and difference variance. These are selected with four different orientations which add up to 16 different parameters. Three statistical methods (raw data analysis, Principle component analysis and non-discriminate analysis) were applied to these texture features. For the classification, the Artificial Neural Network (ANN) and the k-nearest neighbor (K-NN) was used.

Cabezas et al \citep{cabezas2013a}, \citep{cabezas2014b} used BOOST knowledge-based systems to classify MS lesions through voxel. The extracted features include contextual features, outlier maps, and registered atlas-based maps. These features are fed to the Gentleboost classifier. Forty-five different cases were tested to evaluate the proposed work, which has a better performance when compared with previous works. The Boost algorithm showed better performance when the lesion size was tiny. 

The technique in \citep{jesson-a} was used to segment healthy tissues with MS lesions. MRF was used to estimate the tissue label via multi-atlas fusion. A random forest classifier was used to classify these tissues and lesions. The evaluation results showed high accuracy. This method also got the first position in the LMSLSC 2015.

Automatic detection of MS lesions was proposed by Kuwazuru et al \citep{kuwazuru2012a}. This method used T1, T2 and FLAIR images to detect segment lesions by using ANN controlled set method. In the very first step, the lesion background was subtracted, and the lesion was enhanced. Initially, growing regions and multi-gray level thresholding methods were used to detect MS candidates. Fifteen gray-level features and five shapes were used as a feature that include diameter, circularity and the Euclidean distance between the centroid of the brain and the candidates. Mean, variance, and minimum and maximum pixel values were included in the gray level features. SVM was used to classify the MS candidates and the ANN-controlled set method was used to reduce the number of FPs.

Bassem et al \citep{abdullah2012a} proposed the automatic segmentation of MS by using T1 and T2 images. The pre-processing step consists of intensity correction and registration of MRI. Probabilistic MNI Atlas \citep{evans-a} is used to provide each voxel with the probability of belonging to GM, WM or CSF. The images were divided into blocks, and 39 textures features were extracted for each block. To train the SVM, the pixels, which are connected with MS lesions, were labeled manually. SVM used these texture features to classify the MS lesion pixels for both simulated and real datasets. DS, Detected Lesion Load (DLL), TP and TN measures were used for evaluation. 

A Massive Training ANN (MTANN) approach was proposed by Khastavaneh et al \citep{khastavaneh2014a} using T2 and FLAIR images. Brain tissue extraction, bias correction, and intensity normalization were done in the preprocessing step. An image was divided into sub-regions by convolving windows of sizes 13 × 13, 17 × 17 and 21 × 21. Intensity and neighbor voxels were used for feature extraction of each region. The MTANN was trained with these features to classify each voxel as a lesion or as a non-lesion. To reduce the FPs, a fuzzy inference system was used.

Veronese et al \citep{veronese2014a} proposed a method to segment GM lesions by using double inversion recovery (DIR) MR images. These images enhance GM tissues while attenuating WM and CSF which can tribute to the classification of GM lesions. Skull removing, inhomogeneity correction and extraction of brain tissue was done in the pre-processing steps. Initially, the candidate lesion was identified using hyper-intense regions. Local mean, standard deviation, GM around voxel, and neighborhood voxels were used as features fed to the SVM with a radial basis function, to classify GM lesions.

Elliott et al \citep{elliott2014a} proposed an automatic detection method to detect MS lesions using a Bayesian framework via serial MRIs, including T1-w, T2-w, and FLAIR. The study showed a series of scans taken in different weeks showing the incremental lesion process; and detected old lesions, new lesions, and changes in the volume of lesions. The main features considered are lesion intensity, lesion size, and intensity difference. The method was compared with a semi-automatic method to evaluate the performance of the proposed approach. 

Fiorini et al \citep{fiorini2015a} used 91 features from 457 patients, including Patient Report Outcomes, clinical scales, and questionnaires, to build a learning model. The missing data was extracted through the features median values and the dataset was normalized with respect to the minimum and maximum values. PCA was used to remove uncorrelated features and reduce the dimensionality. Different classifiers were used for comparison, including SVM, with high accuracy, and F-score.

Roy et al \citep{roy2015a} developed a longitudinal patch-based segmentation approach to classify WM lesions. 4D patches were extracted from T1-w and FLAIR images, from each subject, and for all time points. A feature dictionary was used to compare the 4D patches from subjects to the 4D patches in the reference dictionary. These features matched in such a way that a dictionary feature can reconstruct the subject patches. These patches are then combined to generate 4D memberships of lesions. The approach was compared with 3D patch-based methods, and DC was used for evaluation.

Loizou et al \citep{loizou2011a} proposed a method based on SVM by using texture features from T2-w MRIs. The patients’ scans were first divided by using the expanded disability status scale (EDSS), EDSS $\leq$ 2 and EDSS $>$ 2. These groups were used in an algorithm that can detect lesions at later stages. Texture features include statistical features, spatial features, and the sum of square variances, moments, correlations and entropy.

Santos et al \citep{santos2016a} proposed a method to substitute  3D metrics with 2D metrics, as 3D metrics are impractical for learning targets and more computational for iterative training, using an oriented training strategy for MS lesion segmentation. A multi-layer perceptron network was used as the segmentation model. The proposed method shows that the orientation method shows better segmentation quality that is comparable to the state-of-the-art approaches.

Ghafoorian et al \citep{ghafoorian2016a} proposed a two-stage approach to differentiate lesions from normal brain tissues. In the pre-processing step, registration, bias field correction, removal of un-cerebral tissue and intensity standardization was done. Two classifiers were trained, one for small WMH $\leq$ 3mm, and the second for large WMHs $>$ 3mm lesions. In the first stage, for each size, an AdaBoost classifier was used with 22 features combined to single WMHs’ likelihood in the second stage to increase the performance. The datasets were used for evaluation from the paper \citep{norden2011a}. 

Guizard et al \citep{guizard2015rotation} used 3D MS lesion segmentation with non-local means (NLM). This method used rotation-invariant and multi-channel for the verity of MS lesions. The proposed method works robustly to detect MS lesions regarding their size, shape, and orientation. For evaluation, clinical datasets were used, and the statistical test showed a strong capability to detect the lesion.

Weiss et al \citep{weiss2013a} used dictionary learning and sparse coding techniques to detect MS lesions. This approach consists of three steps. First, the pre-processing step involved brain masking, extract brain tissues and correction of inhomogeneities. Dictionary learning methods were used to learn features from patches that consist of healthy tissues and a small number of lesions. The error was calculated at each patch, and thresholding was used for final lesion segmentation.  Synthetic and clinical datasets were used to evaluate the algorithm’s performance.

Salem at al \citep{salem-a} extracted deformation and intensity features using T1-w images to train the logistic regression algorithm for classification. The follow-up scans were used of 60 patients where 36 has a new lesion while the 24 scans were without lesions.

Wang et al \citep{wang-a} used Fourier entropy features that were extracted and passed to the multilayer perceptron for classification in The modified parameter-free Jaya algorithm was used for training the multilayer perceptron. A cost-sensitivity learning method was used to overcome the imbalanced dataset problem.

Zhang et al \citep{zhang2019a} detected the conversion of MS from clinically isolated syndrome (CIS). This work used 84 patients diagnosed with CIS and the conversion criteria were define using the 2010 McDonald criteria. The conversion was detection on an average of 82\% using shape features, which contributed to strongly than intensity features.

Han and Hou \cite{han2019a} proposed a method based on feed-forward neural work with adaptive genetic algorithms. The wavelet entropy was used as the main feature with three levels of wavelet decomposition and evaluated via 10-fold cross-validation.

\subsection{Unsupervised Methods}

This section consists of those methods which use unsupervised approaches for classification. These learning techniques are used to classify unlabeled data. These algorithms don’t need any training data or prior information. These techniques include clustering, expectation-maximization and moment’s methods.

A joint histogram-based method was proposed by Zeng et al \citep{zeng2011a} for MS lesion segmentation. Two MR images and T1 and T2 histograms are compared to identify different brain tissues like WM, GM, and MS lesion regions. To detect the lesion, it, firstly, clusters the center coordinate of WM and GM to estimate the lesion regions. The class-adaptive Gaussian Markov modeling approach is used with spatial and density information to classify the joint histogram in different classes. The Fuzzy-C mean clustering is used to identify normal tissues by removing these tissues and leaving the lesion regions. The advantages of this method include robustness to noise and segmentation of small lesions.

Kroon et al \citep{kroon2008a} used a Principle Component Analysis (PCA) based model to classify  MS lesions. Voxel-based features were extracted, including voxel and neighbor voxel intensities, as feature vectors. The distance between the voxel and the feature vector was calculated, including the mean and variance features. PCA was used as a classifier.

Wu et al \citep{wu2006a} used T1-w, T2-w and PD-w images to segment lesions into three different types: enhancing lesions, T1 black holes (BH) and T2 hyperintensities. The Intracranial Cavity (IC) mask was used to extract large vessels, the skull, and extracranial tissues from the brain. Partial artifact correction (TDS+) and template-driven segmentation were tuned to classify MS lesions with the k-nearest neighbor algorithm, KNN. The methods of Wei et al \citep{wei2002a} and Warfield et al \citep{warfield1995a} were used to extract WM via nonlinear registration due to the overlap in tissue. K-NN was used to do multi-classification and to classify BHs.

Ozyavru et al \citep{ozyavru2011a} proposed the fuzzy clustering technique to segment  MS lesions with spatial information. Conventional Fuzzy clustering has been improved by adding features of segmentation elliptically, which is the first step.  The second step improves the algorithm and makes it more robust. Spatial information is merged with intensities based on the standard deviation. The spatial information, based on the Euclidean distance, segments vertical elliptic of brain shapes which define a shape-selective function. The total membership value may be lower than one in this method. If the gray level is more dissimilar to a pixel than multiple standard deviations of average intensity, then that pixel will be filtered out.

Xiang et al \citep{xiang2013a} used a PD-w image and subtracted it from the corresponding T1 to get an enhanced CSF. Using this enhanced image corresponding to T2 images, CSF and MS lesions were extracted by using kernel fuzzy c-mean (KFCM). Thresholding and median filters were used to improve segmentation performance. 

Fartaria et al \citep{fartaria2016a} used four different MRIs: 3DFLAIR, magnetization-prepared rapid gradient echo (MPRAGE), 3D  DIR and magnetization-prepared two inversion-contrast rapid gradient-echo (MP2RAGE), to segment early-stage MS lesions. In the preprocessing step, template atlas registration, tissues’ (GM, WM and CSF) label extraction and intensity standardization were performed. In the classification step, K-NN was used to classify voxels based on their probability. The approach used for K-NN was fuzzy-based rather than a binary K-NN type approach.

Kawa et al \citep{kawa2007a} proposed a fuzzy c-mean (FCM) approach for fast detection. This approach was based on the following steps: first, a Gaussian shape kernel matrix was created with a standard deviation equal to half the standard deviation of the input data. Secondly, two clusters were generated by running the FCM algorithm over the kernelized features. Thirdly, two thresholds and some features in the context of histograms were calculated. The threshold is used to differentiate WM, GM, MS, and their backgrounds. The Gaussian kernels are used for the initial classification of tissues. The classification mask and regions of interest are used in MS lesion detection. 

Shen et al \citep{shan2008a} proposed a method that needs only one anatomical MRI scan and uses intensity-based segmentation by using FCM. This method’s steps include extraction of the intracranial cavity from the original image, using fuzzy membership functions for four different categories, and using the Gaussian kernel to smooth these membership functions. Probability maps were calculated using SPM5. Inconsistencies were calculated between pixel intensities. The classification of the lesion was based on the inconsistencies. Normalization and morphological operation are used in the post-processing step. However, the method didn’t detect lesions less than 1 $cm^3$.

Horsefiled et al \citep{horsfield2007a} proposed a semi-automatic segmentation method based on fuzzy connectedness. The  ROI is extracted from PD-w images and used as a binary mask. The new transformed images are down-sampled with 64 × 64 × 64 pixels, to give it an anatomical position. The correlation is calculated for each pixel from the corresponding binary image. The 3D kernel on each pixel is used to find the conditional probability of lesions on each pixel. To estimate the lesion size, an exponential fit is used to build a template image. These template images are used in fuzzy connectedness and start with seed pixels to find ROIs and draw boundaries around the lesions.

Washimkar and Chede \citep{washimkar2015a} used a saliency map approach to detect the region of interest in an MR image. To segment abnormal tissues, the AM-FM approach is used in MRI images. FCM clustering and saliency mapping are used to classify the infected regions in MRI images.

Bhanumurthy and Anne \citep{m2016a} proposed a modified algorithm based on the Histon fast fuzzy C-means (HFFCM). Energy, correlation, entropy and shape features were used and compared with other methods for evaluation. The MICCAI dataset was used for these experiments.

Ong et al \citep{ong2012a} proposed an automatic segmentation method for WM lesion detection. This method used an adoptive outlier detection approach by using FLAIR images. This method used skull stripping and inhomogeneity correction in the pre-processing step. The threshold point was determined adaptively to segment and detect WM lesions. A histogram of the skull-stripped image was constructed, and a Gaussian kernel was applied to smoothen the histogram, removing the artifacts e.g. noise, which affect the outlier detection accuracy.

Zeng et al \citep{zeng2015a} proposed an unsupervised algorithm based on a joint histogram modeling. First, the histogram was generated for T1, T2 and FLAIR images to identify the lesion areas and then to convert it back to a 2D image. The prior information, used to reduce the false positives, was followed by alpha matting to overcome the tissue density overlapping problem.

Jain et al \citep{jain2015a} worked on MSmetrix which is a reliable and automatic tool for MS lesion segmentation. 3D T1 and FLAIR used a probabilistic model to detect WM as outliers with respect to GM. The real MS lesion will be extracted based on some prior knowledge, including location and appearance. 

Valverde et al \citep{valverde2014a} proposed a WM lesion filling technique to find tissue volume. The voxel intensity was replaced by the mean of the WM intensity of each of the two-dimensional slices which achieve better performance. The WM lesion is first registered and then its voxel intensity is replaced between WM and GM tissues. To find the volume of the tissues, Independent tools, FAST and SPM8, were used which, compared to other state-of-the-art methods, performed well. The proposed method is also used as an extension of the SPM tool.

Steenwijk et al \citep{steenwijk2013a} proposed the k-nearest neighbor (KNN) with tissue type priors-based, TTPs, method for automatic segmentation of WM lesions by using 3DFLAIR and 3DT1 images. In the pre-processing step, the brain was extracted, and the non-brain tissues were removed, followed by the inhomogeneity correction step. The features include 3DFLAIR intensities and 3DTI intensities for the training set, and voxels were labeled manually. These features were normalized with robust range normalization \citep{boer2009a} and histogram matching \citep{younis2008a} because these methods are robust to intensities. The KNN method was used as a classifier. The post-processing was applied to reduce the number of false positives by removing lesions that have a smaller size than some threshold.

Jerman et al \citep{jerman2016a} proposed a hybrid of supervised and unsupervised methods to segment MS lesions. The method used an unsupervised mixture model to extract the lesion that was classified via random decision forest (RDF) with neighborhood voxel intensities. The RDF used a balanced training set for lesions and non-lesion voxels. It also used visual features for discrimination which is easily extracted from MR images to avoid the computation overhead.

Bonanno et al \citep{bonanno2021multiple} developed an automatic CAD tool for the purpose of relapsing remitting MS lesion classification. The proposed method is to use watershed transformation for segmenting lesions and extracting features. Afterwards, city block norms clustering analysis is applied to group lesions and identify suspicious ones. The input is a T2 FLAIR image sequence. The tool was evaluated on a dataset of 20 MRI images. The result achieved was  87\% diagnostic accuracy, i.e. specificity,  on AUC and 77\% on the sensitivity score.

\subsection{Deep Learning Methods}

Brosch et al \citep{brosch2015a} proposed a method based on a deep convolutional neural network. The convolutional neural network uses the entire MRI scan instead of various patches, learns features by itself, and classifies the lesions. The architecture of the network consists of three main layers which include an input layer that takes the MRI, a convolutional layer with 32 filters through which the network can learn features and a deconvolution layer that uses the learned features to classify lesions. The objective function was modified in such a way that it can deal with unbalanced classes. 

Yoo et al \citep{yoo2014a} presented an automatic learning approach by using unlabeled images to segment MS lesions. A deep learning approach was used to extract features from unlabeled data while some of the images were labeled for supervised learning. A large amount of data was used for training; 1400 cases were used in which 100 were labeled cases. Two-layered Restricted Boltzmann Machines (RBMs) were used. RBMs were stacked together to form a deep belief network to extract a feature vector and feed it to the random forest classifier which was tested with 50 MRIs.

Brosch et al \citep{brosch2016a} used a 3D deep neural network approach which consists of two pathways including a convolutional pathway, which extracts high-level features, and a de-convolutional pathway which estimates the segmentation. These two pathways provide high and low-level features that can be integrated to achieve better segmentation across a various range of lesion sizes. A deep 3D convolutional encoder network was applied on MR images collected from two different datasets, MICCAI 2008 \citep{styner2008a} and the ISBI 2015 challenge \citep{petrick2013evaluation}.

Birenbaum and Greenspan \citep{birenbaum2016a} used an applied Convolutional Neural Networks (CNN) to predict  MS lesions. The approach involved a pre-processing phase that consists of various operations including co-registration, bias field correction, brain extraction and intensity normalization. In the second phase, a candidate lesion, acting as prior knowledge, was extracted via a mask. In the third phase, a multi-view convolutional neural network is used to predict the lesion probability. 

Valverde et al \citep{valverde2017b} proposed a method based on two 3D-patch-wise CNN. The first network was trained to extract the lesion from MR images while the second network was used to reduce the number of misclassified voxels as a result of the first network. The method was evaluated with MICCAI 2008 \citep{styner2008a} and two MS private clinical datasets.

Prieto et al \citep{prieto-a} used  CNN with a large data sample of 900 subjects which were extracted from The Partners of Comprehensive Longitudinal Investigation of MS (CLIMB) repository \citep{unknown-h}. The convolutional network consists of 9 layers including 2 convolution and pooling layers with 5 fully connected layers.

Another study \citep{zhang2018a} used 10-layer CNN with dropout and parametric ReLU layers. These layers will help CNN to be more generalized. Two different datasets were used, eHealth dataset and a private dataset, which consist of 38 and 26 individuals and with slices of 676 and 681, respectively. As the deep learning models need more data for training, the authors adopted the data augmentation approach to increase the data and achieve better performance in terms of accuracy and sensitivity. 

Roy et al \citep{roy2018a} used fully CNN using multi-contrast MRI to segment WM lesions. This approach used two pathways, the first approach used a various filter on MR images and the output will be combined to the second part for further filters. It will produce a member function to make a binary mask for lesions segmentation. 

Valverde et al \citep{valverde2019a} used the pre-trained CNN model which was trained on two datasets, MICCAI 2008 and MICCAI 2016. It will reduce the new training and test data, as the model already tuned on previous weights. This approach was evaluated on two datasets and improved the detection rate with a small number of training images. 

Wang et al \citep{wang-b} used 14-layer CNN with dropout, stochastic pooling and normalization steps to identify MS in brain images. The data augmentation step was used to improve CNN performance. This approach used two datasets for evaluation and achieved better performance. 

The work in \citep{nair2020a} discussed the uncertainty of DL methods and used a 3D CNN for segmentation and detection of MS lesion. Two different approaches were used: voxel-wise and lesion-wise thresholding for detection lesions. A prediction variance, Monte Carlo sample variance, predictive entropy and mutual information metrics were used for evaluation.

Aslani et al \citep{aslani2019a} used multi-modal brain images to segment MS lesions. The end-to-end 2D CNN was used for slice-based segmentation. The downsampling path was used to extract information separately from multiple modalities and upward sampling was used to combine features maps to extract lesion location and shapes. A private dataset of 37 MS patients and a public dataset ISBI 2015 was used for training and testing.

Salem et al \citep{salem-b} used the augmentation method to improve encoder and decoder U-Network (U-Net) performance. The synthetic lesions were generated to increase the training data for the two-input and two-output convolutional algorithm. The proposed method was evaluated on two datasets a private and ISBI 2015 and showed the performance of the convolution network via synthetic data augmentation. 

Hashemi et al \citep{hashemi-a} worked on a 3D-FC-dense network with large overlapping patches as an input image. The overlapping patch image was used for data augmentation, patch selection, and patch prediction fusion via B-spline weighted soft voting for the prediction at patch border. The proposed method was evaluated using two datasets MSSEG and ISIB with the performance metrics of focal loss, generalized dice loss (GDL) and $F_\beta$ loss. 

Gabr et al \citep{gabr2019a} used a fully convolutional neural network (FCNN) for the classification of MS lesions. The proposed method was evaluated around 1000 MRIs of RRMS patients with cross-validated and classify WM, GM CF and T2 MS lesions. 

Feng et al \citep{feng2019a} achieved the 2nd rank in IEEE ISBI 2015 longitudinal lesion challenge dataset using a 3D-U-Net network with dropout layers for segmentation and achieved the DSC or 0.684.
Narayana et al \citep{narayana2018a} worked on CNN using T1-w and FLAIR images to detect MS lesions. The pre-processing was used for skull stripping, co-registration, bias field correction, denoising and intensity normalization. 

Zhang et al \citep{zhang2018b} used a generative adversarial network (GAN) for MS lesion localization. The pipeline used an encoder-decoder generator and various discriminators corresponding to multiple inputs modalities.

Li et al \citep{li2019a} proposed a multi-scale aggregation model to detect small and large lesions. Stack-Net was designed to capture local features to segment small lesions and multi-scales Stack-Nets were combined with different receptive fields to capture multi-scale contextual features for large lesions. 

Aslani et al \citep{aslani2019b} used encoder-decoder CNN, where encoder extracts the features via different features maps and the decoder combines the features map via up-sampling. This was the slice-based CNN for MS lesion segmentation and was evaluated via ISBI MS challenge dataset. 

Sepahvand et al \citep{sepahvand2019a} also used 3D-CNN which was training via a large amount of MR images labeled by experts. CNN was used for prediction and the modified U-Net was used for segmentation purposes.

Salem et al \citep{salem2020a} used FCNN with longitudinal brain MRI to identify new T2-w lesions. Two U-net blocks were used, first to learn deformation fields and the second was used to segment the final lesions. The model used end-to-end learning for both deformation fields and new lesions. In segmentation, the average DSC of 0.55 was achieved.

Mckinley et al \citep{mckinley2020a} proposed the DeepScan MS approach to segment progressive MS lesions. This work separated stable patients from progressive patients via lesion count and lesion volume. Three private datasets were used to evaluate the proposed deep learning approach.

Rosa et al \citep{rosa2019a} compared the shallow and deep neural networks and investigated their performance via MS lesion segmentation. K-NN and CNN were used for segmentation separately and jointly for training and validation. The combined approach performed better than the standalone approach.

Sepahvand et al \citep{sepahvand2020cnn} introduced a U-Net based deep CNN classification method to detect New and Enlarging lesions (NE Lesions) of  relapsing remitting MS. They improved the segmentation by subtracting images along with different time points then element-wise multiplying these images with the base references. They compared their method with three different approaches including DeepMedic architecture.

Ye et al \citep{ye2020deep} approached lesions classification by employing a deep neural network of 10 hidden layers. They, previously, developed a diffusion basis spectrum imaging DBSI technique where they employed it in this research as a feature extractor along with MRIs intensities. 

Kamraoui et al \citep{kamraoui2020towards} proposed an ensemble of CNNs. The ensemble helps in reducing generalization errors of its constituting networks. They also synthesized multiple instances on-the-fly with various filters of images in the dataset to circumvent disparity in images quality. Another contribution they proposed is a hierarchical specialization learning where they initialized the model from pretrained weights instead of random initialization. They evaluated the model on ISBI and MSSEG’16 datasets and a third in-house dataset. 

Gessert et al \citep{gessert2020multiple} proposed the problem of new and enlarged lesions segmentation by developing a two-path encoder-decoder attention-guided fully-convolutional CNN model based on U-Net. The two paths process indicates that the image is processed individually on the encoder level then jointly on the decoder level. They evaluated the model on two datasets, ZURICH dataset of 89 MS cases and another one of 97 labeled pairs from 33 patients. They experimented their method with other methods existing in the literature and with a single-path architecture where two paths architecture proves superiority on Dice TPR and FPR metrics. 

La Rosa et al \citep{la2020multiple} introduced a fully convolutional CNN to detect MS cortical lesions. The proposed model is based on 3D-U-Net. The approach is evaluated on a dataset of 90 patients from two different centers. They consist of 728 and 3856 GM and WM lesions respectively. The achieved detection rate reached 76\% for both WM and GM lesions and false positive rate of 29\%.

Krüger et al \citep{kruger2020fully} devised an encoder-decoder convolutional CNN based on U-Net architecture. The network is evaluated against LST toolkit. The results show that the author's architecture outperformed LST on various measures. The evaluation was conducted on Zurich and Dresden datasets. 

\subsection{Fuzzy Methods}

Fuzzy approaches are those which are based on truth values rather than 0s and 1s. This idea was introduced by Zadeh \citep{zadeh1988a}. It contains membership functions and fuzzy rules to classify objects. In automatic MS lesion segmentation, there are some challenges like sensitivity, which may reduce false positives. One of the main reasons for this is to reduce the rate of small lesion detection. Because small lesions give large weight to the boundary pixel, false detections increase. To overcome false positives and increase sensitivity, the Mamdani type fuzzy rule-based system \citep{aymerich2011a} is designed to detect MS lesions increasing the sensitivity level in the presence of small lesions. The main four steps of the proposed algorithm included to achieve the goal are pre-processing, fuzzy characterization (with the evaluation of hyperintensity characteristics), defuzzification and filtering of false detections.

A fully automatic segmentation method was proposed by Admiraal-Behloul \citep{admiraal-behloul2005a} to extract the WMM volume in elder patients by using T2-w, PD-w and FLAIR images.  CR-FLAIR \citep{woods1998a} is used for the registration of PD-w and FLAIR images. An atlas provides a 3D probability model and prior probability for GM, WM and CSF. FCM is used to classify PD-w image as a background and as a foreground. The other two type T1-w and CR-FLAIR are classified into three classes including, dark, bright and medium bright. The fuzzy membership function is used to classify CSF and WMH.

Bijar et al \citep{bijar2012a} proposed a fuzzy membership automatic segmentation approach for MS lesions. They partitioned the brain into three parts, dark region CSF and gray region, where each region has its membership function maximizing the fuzzy entropy which is done via GA. The images obtained via the fuzzy membership functions are classified into brain tissues. Through this three-level threshold, the segmentation is not done with good manners because the membership function defines an MS lesion that has a higher intensity than CSF, but, in some cases, they are similar to the CSF, due to which they belong to the same class. This problem which is due to noise is tackled by applying a localized neighborhood filter to the high-intensity class.

Esposito et al \citep{esposito2010a} used a fuzzy evolutionary approach that defines a fuzzy inference technique to classify the WML while the evolutionary algorithm tunes the membership function for that variable which is in the involved rules. According to the author, it is the first time in literature that the adaptive fuzzy approach is proposed for such type of problems. During the experiments in the first generations, the accuracy was 30-35\% while in the upper generation, the accuracy increased rapidly: up to 85-90\% on average.

The work in \citep{aymerich2011a} has been extended in \citep{aymerich2011b} to focus more on small lesion detection and to apply different filters to reduce false positives.  Hyperintensities were used to detect lesions in the 9 × 9 windows. The fuzzy set was used to evaluate the pixel intensities in T2-w images. Mamdani type fuzzy rule-based system (MFRBS) was used, which consider usually logical AND operators. In the end, the binary image was generated for neuro-radiologist evaluation. This work used 148 Proton Density (PD) and T2-w MRIs to evaluate the algorithm. The results showed that the proposed algorithm reduces the false positives with up to 3.6\%.

Esposito and Pieto \citep{esposito2011a} proposed a work based on an ontology-based fuzzy decision support system (DSS) which classifies WML and computes their volume. The algorithm was evaluated on 162 patients’ data. Thresholding was applied to fuzzy outputs to obtain the binary classification and then evaluated by a ROC curve to show the tradeoff between sensitivity and specificity. Statistical analysis has been done to check the influence in the diagnosis of MS lesions.

Ghahazi et al \citep{ghahazi2014a} proposed a fuzzy rule-based approach in which age, score and the possibility of being diagnosed are the membership function for input. The rule was used to evaluate algorithm performance.

Khastavaneh et al \citep{khastavaneh2014b} proposed a method to reduce the number of FPs using a fuzzy inference system. The method consists of a preprocessing step which includes skull removing, brain tissue extraction, registration and normalization of inhomogeneities. In the fuzzification step, four different parameters were used including initial lesion, voxel intensity, neighboring intensities, and lesion likelihood. In the inference step, four membership functions were used based on the previous parameters with low, medium and height (high) membership values. The defuzzification was used as the last step to improve the results.

De et al \citep{falco2016a} proposed a decision-based method based on differential evolution. This approach automatically extracts the explicit knowledge from the MS database via defining their own set of ‘IF-THEN’ rules which are concatenated through ‘AND’ with database attributes.

\small
\begin{table*}[]
\centering
\resizebox{\textwidth}{!}{%
\begin{tabular}{|c|c|c|p{3.5cm}|p{3.75cm}|c|}
\hline
\textbf{Type} &
  \textbf{Study} &
  \textbf{Year} &
  \textbf{Algorithms} &
  \textbf{Image Sequence} &
  \textbf{Lesions} \\ \hline
\multirow{12}{*}{Data-driven} &
  Liu et al \citep{liu2009a} &
  2009 &
  Texture feature &
  T2-w &
  WML \\ \cline{2-6} 
 &
  Cebezas et al \citep{cabezas2014a} &
  2014 &
  MEM &
  T1, T2, PD, FLAIR &
  MSL \\ \cline{2-6} 
 &
  Villalon et al \citep{villalon_reina2013a} &
  2013 &
  BAVS &
  FLAIR &
  MSL \\ \cline{2-6} 
 &
  Hill et al \citep{hill2014a} &
  2014 &
  IJM &
  Seudo-Color MRI &
  MSL \\ \cline{2-6} 
 &
  Schmidt et al \citep{schmidt2012a} &
  2012 &
  VBM &
  T1-w, FLAIR &
  HL \\ \cline{2-6} 
 &
  Wack et al \citep{wack2013a} &
  2013 &
  MACC &
  FLAIR &
  HL \\ \cline{2-6} 
 &
  Zhong et al \citep{zhong2014a} &
  2014 &
  Thresholding + Region Growing &
  FLAIR &
  WML \\ \cline{2-6} 
 &
  Ganiler et al \citep{ganiler2014a} &
  2014 &
  Thresholding &
  T2-w, T2-w, PD-w &
  WML \\ \cline{2-6} 
 &
  Wiggermann et al \citep{wiggermann2016flair2} &
  2016 &
  SNR + CNR &
  T2, FLAIR2 ­, FLAIR &
  WML \\ \cline{2-6} 
 &
  Storelli et al \citep{storelli2016a} &
  2016 &
  Region Growing &
  T2, DEPD &
  MSL \\ \cline{2-6} 
 &
  Cabezas et al \citep{cabezas2016a} &
  2016 &
  IR + DF &
  T1-w, T2-2, PD-2 &
  T2-L \\ \cline{2-6} 
 &
  Rodrigo et al {[}74{]} &
  2015 &
  Pixel intensity feaures &
  FLAIR &
  MSL \\ \hline
\multirow{14}{*}{Statistical} &
  Khayati et al \citep{khayati2008a} &
  2008 &
  Bayesian &
  FLAIR &
  MSL \\ \cline{2-6} 
 &
  Harmouche et al \citep{harmouche2015a} &
  2015 &
  MRF &
  T1, T2, PD, FLAIR &
  T1-HL, T2-HL \\ \cline{2-6} 
 &
  Shiee et al \citep{shiee2010a} &
  2010 &
  FCM &
  T1, T2, PD &
  WHL \\ \cline{2-6} 
 &
  Harmouche et al \citep{harmouche2006a} &
  2006 &
  Bayesian + MRF &
  T2-w, PD-w &
  T2 HL \\ \cline{2-6} 
 &
  Bricq et al \citep{bricq2008a} &
  2008 &
  HMM + TLE &
  T1-w, FLAIR &
  MSL \\ \cline{2-6} 
 &
  Prastawa et al \citep{m2008a} &
  2008 &
  PDF + MCD &
  T1-w, T2-w, FLAIR &
  MSL \\ \cline{2-6} 
 &
  Aït-Ali et al \citep{a2005a} &
  2005 &
  TLE &
  T1-w, T2-w PD-w &
  T1-HL, T2-HL \\ \cline{2-6} 
 &
  Gao et al \citep{gao2013a} &
  2013 &
  EnM &
  T1-w, T2-w, FLAIR &
  WML \\ \cline{2-6} 
 &
  Gong et al \citep{gong2015a} &
  2015 &
  EnM &
  FLAIR &
  MSL \\ \cline{2-6} 
 &
  Tomas-Fernandez et al \citep{tomas-fernandez2015a} &
  2015 &
  GMM &
  T1, T2, FLAIR &
  MSL \\ \cline{2-6} 
 &
  Elliott et al \citep{elliott2013a} &
  2013 &
  Bayesian + RF &
  FLAIR &
  MSL \\ \cline{2-6} 
 &
  Gao et al \citep{gao2014a} &
  2014 &
  EnM + BFE &
  T1-w, T2-w, FLAIR &
  MSL \\ \cline{2-6} 
 &
  Zangeneh et al \citep{zangeneh2016a} &
  2016 &
  GMM + ANN &
  T1-w, T2-w, FLAIR &
  MSL \\ \cline{2-6} 
 &
  Zhao et al \citep{zhao2017a} &
  2017 &
  EnM &
  T1-w, FLAIR &
  MSL \\ \hline
\multirow{11}{*}{Supervised} &
  Roy et al \citep{roy2013a} &
  2013 &
  SVM &
  T1-w, T2-w, FLAIR &
  MSL \\ \cline{2-6} 
 &
  Deshpande et al \citep{deshpande2015a} &
  2015 &
  ADL &
  T1-w MPRAGE, T2-w, PD, FLAIR &
  MSL, WM, GM Tissues \\ \cline{2-6} 
 &
  Zhang et al \citep{zhang2007a} &
  2007 &
  ANN + KNN &
  T2-w &
  \multicolumn{1}{l|}{MSL, WM Tissues} \\ \cline{2-6} 
 &
  Cabezas et al \citep{cabezas2013a} &
  2013 &
  Gentleboost Algorithm &
  PD-w, T2-w &
  HL \\ \cline{2-6} 
 &
  Cabezas et al \citep{cabezas2014b} &
  2014 &
  Gentleboost Algorithm &
  T1-w, T2-w, PD-w, FLAIR &
  MSL \\ \cline{2-6} 
 &
  Guizard et al \citep{guizard2015rotation} &
  2015 &
  RMNMS &
  T1-w, T2-w, FLAIR &
  MSL \\ \cline{2-6} 
 &
  Jesson et al \citep{jesson-a} &
  2015 &
  MRF + RF &
  T1, T2, FLAIR &
  GML \\ \cline{2-6} 
 &
  Kuwazuru et al \citep{kuwazuru2012a} &
  2012 &
  ANN + levelSet method &
  T1, T2, FLAIR &
  MSL \\ \cline{2-6} 
 &
  Abdullah et al \citep{abdullah2012a} &
  2012 &
  SVM &
  T1, T2 &
  MSL \\ \cline{2-6} 
 &
  Khastavaneh et al \citep{khastavaneh2014a} &
  2015 &
  MTANN + FIS &
  T1-w, T2-w, FLAIR &
  HL \\ \cline{2-6} 
 &
  Veronese et al \citep{veronese2014a} &
  2014 &
  SVM &
  DIR &
  GML \\ \hline
\multirow{12}{*}{Unsupervised} &
  Wu et al \citep{wu2006a} &
  2006 &
  KNN &
  T1-w, T2, PD-w &
  T2-HL \\ \cline{2-6} 
 &
  Kawa et al \citep{kawa2007a} &
  2007 &
  FCM &
  T1, FLAIR &
  MSL \\ \cline{2-6} 
 &
  Kroon et al \citep{kroon2008a} &
  2008 &
  PCA &
  T1, T2, FLAIR &
  MSL \\ \cline{2-6} 
 &
  Ozyavru et al \citep{ozyavru2011a} &
  2011 &
  FC &
  T1, T2 &
  MSL \\ \cline{2-6} 
 &
  Bhanumurthy et al \citep{m2016a} &
  2016 &
  HFFCM &
  T1, T2, FLAIR &
  MSL \\ \cline{2-6} 
 &
  Xiang et al \citep{xiang2013b} &
  2013 &
  FCM &
  T1-w, T2-w, PD-w &
  WML \\ \cline{2-6} 
 &
  Jain et al \citep{jain2015a} &
  2015 &
  MSmetrix &
  T1-w, FLAIR &
  HL \\ \cline{2-6} 
 &
  Valverde et al \citep{valverde2014a} &
  2014 &
  FCM &
  T1-w &
  Tissue volume \\ \cline{2-6} 
 &
  Washimkar et al \citep{washimkar2015a} &
  2015 &
  FCM &
  FLAIR &
  Infected regions \\ \cline{2-6} 
 &
  Fartari et al \citep{fartaria2016a} &
  2017 &
  KNN &
  MPRAGE, MP2RAGE; DIR &
  WML, Cortical Lesion \\ \cline{2-6} 
 &
  Zeng et al \citep{zeng2015a} &
  2015 &
  JHM &
  T1, T2, FLAIR &
  MSL \\ \hline
\multirow{5}{*}{Fuzzy} &
  Bijar et al \citep{bijar2012a} &
  2012 &
  Fuzzy Entropy + GA &
  FLAIR &
  MSL \\ \cline{2-6} 
 &
  Esposito et al \citep{esposito2010a} &
  2010 &
  Fuzzy Rules based &
   &
  WML \\ \cline{2-6} 
 &
  Aymerich et al \citep{aymerich2011b} &
  2011 &
  Fuzzy Rules based &
  T2-w &
  MSL \\ \cline{2-6} 
 &
  Esposito et al \citep{esposito2011a} &
  2011 &
  Ontology-based fuzzy decision support (OB-FDS) &
  T1, T2, PD-w, QMCI &
  WML \\ \cline{2-6} 
 &
  Khastavaneh et al \citep{khastavaneh2014b} &
  2014 &
  Fuzzy Inference System &
  FLAIR &
  MSL \\ \hline
\end{tabular}%
}
\caption{Methods categorization with image sequences}
\label{tab:table-img-seq}
\end{table*}

\normalsize

\begin{table*}[]
\centering
\def\arraystretch{1.5}

\begin{tabular}{|c|p{5cm}|p{4.3cm}|p{4.5cm}|}
\hline
\textbf{Types} &
  \textbf{Description} &
  \textbf{Advantages} &
  \textbf{Disadvantages} \\ \hline
Data-driven &
  Used spatial and histogram   information, thresholding, Voxel-based Image processing &
  Consistent overtime &
  Highly sensitive to noise \\ \hline
Statistical &
  These methods used   Probabilistic modeling &
  Can work with few   sample data &
  Sensitive to noise and data distribution \\ \hline
\multirow{2}{*}{Supervised} & Supervised based on   Atlas, which can use as previous information. & Include local   information & Highly data dependent.  \\ \cline{2-4} 
                            & Supervised based on   training, which can be trained using datasets & No need for   registration  & Need enough samples   for training \\ \hline
Deep learning &
  A complex deep neural   network &
  Better classification   accuracy &
  A huge amount of training   data needed \\ \hline
Unsupervised &
  Do not need training   samples &
  No need for labeled   samples &
  Depends on the   quality of tissues \\ \hline
Fuzzy &
  Rule-based   classification & Fuzzy rules are robust, simple, and refined according to requirements
   &
  For complex problem needs more rules. It is also highly dependent on human knowledge and expertise \\ \hline
\end{tabular}%

\caption{Pros and cons of categories}
\label{tab:pro-cons}
\end{table*}

\section{Discussion}
\label{sec:8}
In this work, we reviewed various studies proposing MS CAD systems that detect and segment MS lesions from brain MRIs.  All these studies used various medical image modularities where FLAIR is the best in detecting changes and new MS lesions. Many of these studies also show the follow-up MS lesion (new lesion) detections, such as \citep{cabezas2016a}, \citep{shiee2010a}, \citep{lesjak2016a} and \citep{salem-a}.

Lesion classification can be divided into two main categories: class-based and outlier-based classifications. In a class-based classification, the lesions are grouped along with WM, GM and CSF while in an outlier-based method, the WM, GM and CSF are considered as three different groups and lesions are classified as outliers, also called outlier segmentation. Due to a small number of voxels in a lesion area, the probability density function (PDF) will not be accurate as other classes. Due to this, the outlier-based method performs better segmentation. However, only few studies have been published in this area \citep{m2008a}, \citep{a2005a} and \citep{garc2008a}. 

\begin{figure}
    \centering
    \includegraphics[scale=0.3]{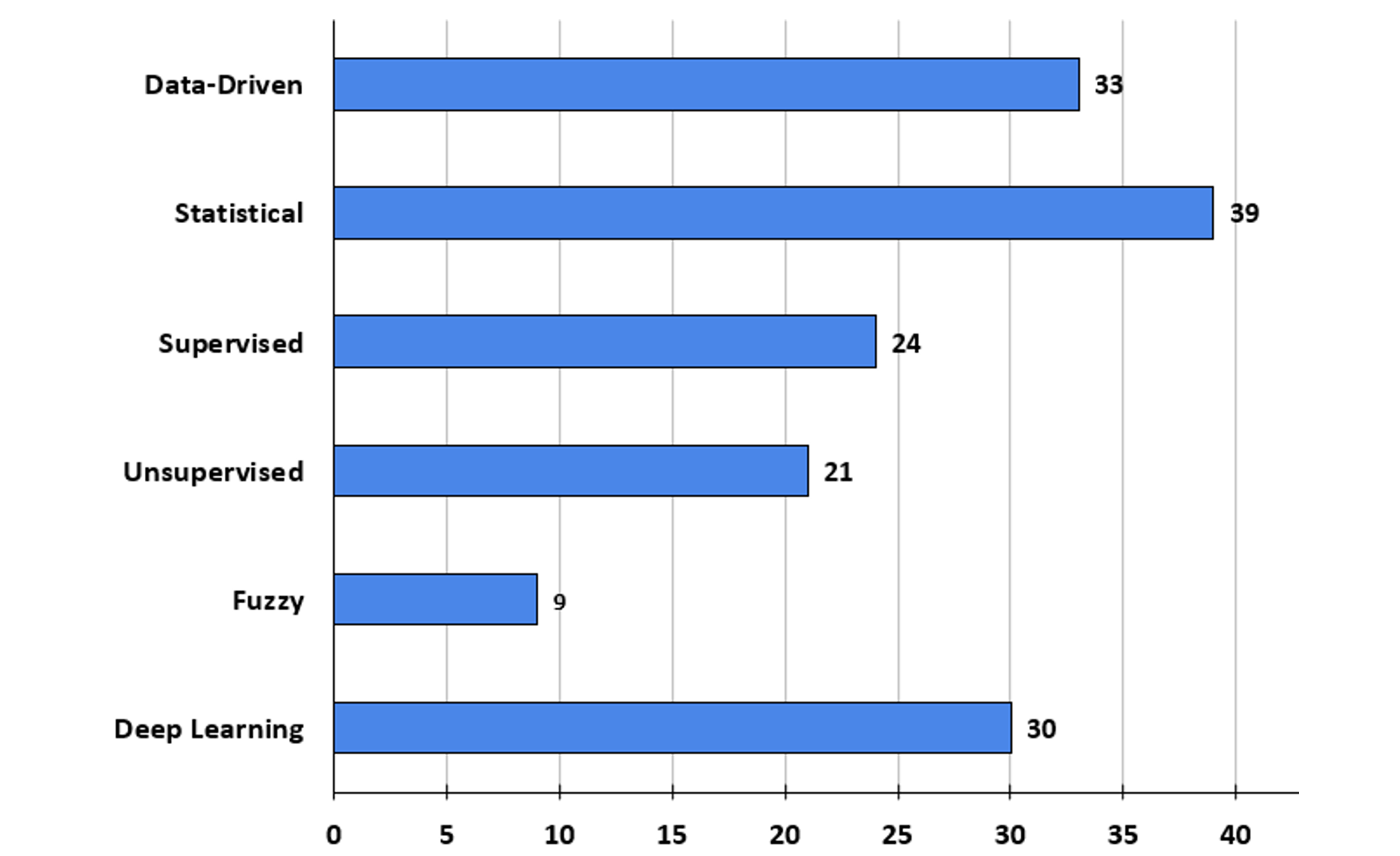}
    \caption{Papers distribution based on methods}
    \label{fig:paper-dist}
\end{figure}

A large number of data-driven based methods are reviewed in this study as shown in Figure \ref{fig:paper-dist}. These methods used thresholding approach which does not consider intensity overlaps of different tissues, which makes it difficult to draw an optimal threshold between the tissue boundaries.  Other than thresholding, a region growing approach is used, where similar voxels grouped together based on their intensities.However, one problem with this approach is that it is sensitive to noise and choosing the seed point is a difficult task because it can vary among tissues. Histogram matching and normalization are also used under the same umbrella where some researchers used GA to optimize the histograms \citep{stamile2017a}.

According to Figure \ref{fig:paper-dist} most of the reviewed papers focus on statistical methods. Recall that statistical methods are the ones that use probabilistic approaches, while those methods which use KNN and fuzzy clustering methods are classified as unsupervised learning.  These methods used non-parametric methods such as KNN and Perzen window for tissue classifications. KNN is a widely used method, however, due to its computation time, it may affect the training process. Statistical methods focus on probabilistic atlases and EM methods. In EM methods, it is assumed that data is a normal distribution which may not be the case due to the variation of intensities in MRIs and can lead to inaccurate results, specifically in brain lesion classification.

Supervised methods were also used in lesion and tissue segmentation using MRIs. These methods used features to train the learning algorithms which include SVM, ANN, dictionary learning and decision trees. These algorithms used various parameters to tune the algorithms for better performance. One of the issues with such algorithms is excessive training time to learn the features i.e., the deep neural networks will take much time due to the many layers and a large number of neurons as compared to shallow neural networks. Such large networks are data hungry and need large amount of data for training.

To avoid learning algorithms, clustering algorithms are used i.e., fuzzy C-means clustering \citep{xiang2013a}. One of the disadvantages of these algorithms is their sensitivity to noise which can be overcome with the use of statistical methods including statistical atlases \citep{shan2008a} and probabilistic maps \citep{shiee2010a}. Some papers also used fuzzy rules-based methods \citep{esposito2010a}, \citep{aymerich2011b}, fuzzy entropy \citep{khastavaneh2014b} and other fuzzy-based methods \citep{khastavaneh2014b}, \citep{esposito2011a}, \citep{bazin2008a}.

Deep learning gained popularity after a breakthrough when Krizhevsky et al introduced AlexNet \citep{krizhevsky2012a} and used deep CNN for the ImageNet challenge. Besides that, other deep learning models including LeNet \citep{lecun1998a}, ZF Net \citep{zeiler2013a}, GoogleNet \citep{szegedy2014a} by Google, VGGNet \citep{simonyan2014a} and ResNet \citep{he2015a} are  popular models. These models are an improved version of ANN which consists of more layers with a higher level of abstraction \citep{lecun2015a}. Training algorithms such as deep networks i.e., CNN, from scratch is a challenging task. First, CNN needs a large amount of training data which may be difficult in the medical domain. Second, the training process needs large memory and computational resources. Third, these models have over fitting and convergence issues which require a repetition of experiments and tuning hyper-parameters. To avoid such issues, transfer learning \citep{shin2016a} and parameter fine-tuning are popular methods. 
Brosch et al \citep{brosch2016a} used the combination of the convolutional pathway, which learned high-level features, and the de-convolutional pathway, which predicts the classes. This approach was compared to five other methods and its performance is considered as state-of-the-art in MS lesion segmentation. Deep learning is not only limited to lesion classification as it is also widely used in medical imaging \citep{greenspan2016a} like brain tumor segmentation \citep{pereira2016a}, lung diseases \citep{anthimopoulos2016a}, nuclei classification \citep{sirinukunwattana2016a} and candidate’s lesion detection \citep{roth2016a},\citep{setio2016a}.

\begin{figure}
    \centering
    \includegraphics[scale=0.3]{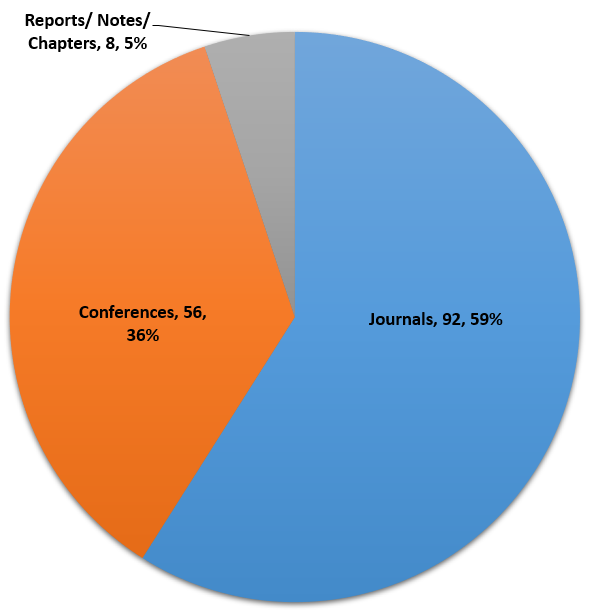}
    \caption{Over all papers distribution}
    \label{fig:overall-dist}
\end{figure}

The analysis shows that a large number of papers are published in the area of MS segmentation and detection. The most widely used methods are statistical, data driven, and deep learning methods while the least used methods are fuzzy based methods, as shown in Figure  \ref{fig:paper-dist}. Further analysis shows that 58\% of the review papers are journal,s 37\% are conferences, and the rest of 5\% are reports/notes/book chapters as shown in Figure \ref{fig:overall-dist}.

\begin{figure*}
    \centering
    \includegraphics[scale=0.55]{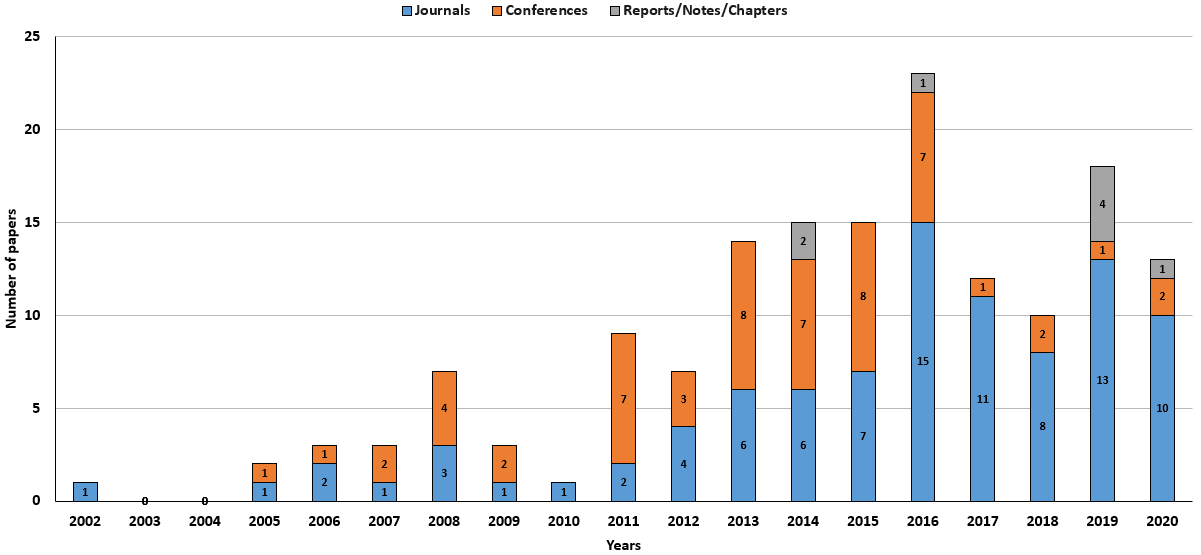}
    \caption{Year-wise distribution of journals, conferences and reports.}
    \label{fig:yw-num-papers}
\end{figure*}

To see the yearly-wise published papers and their types, Figure \ref{fig:yw-num-papers} shown some interesting results. Initially, MS segmentation and detection methods get published both in conferences and journals; however, after 2016, the number of journal papers increased significantly and the number of conference papers reduce rapidly while only a few reports, lecture notes and book chapters are published in 2019.

\begin{figure*}
    \centering
    \includegraphics[scale=0.30]{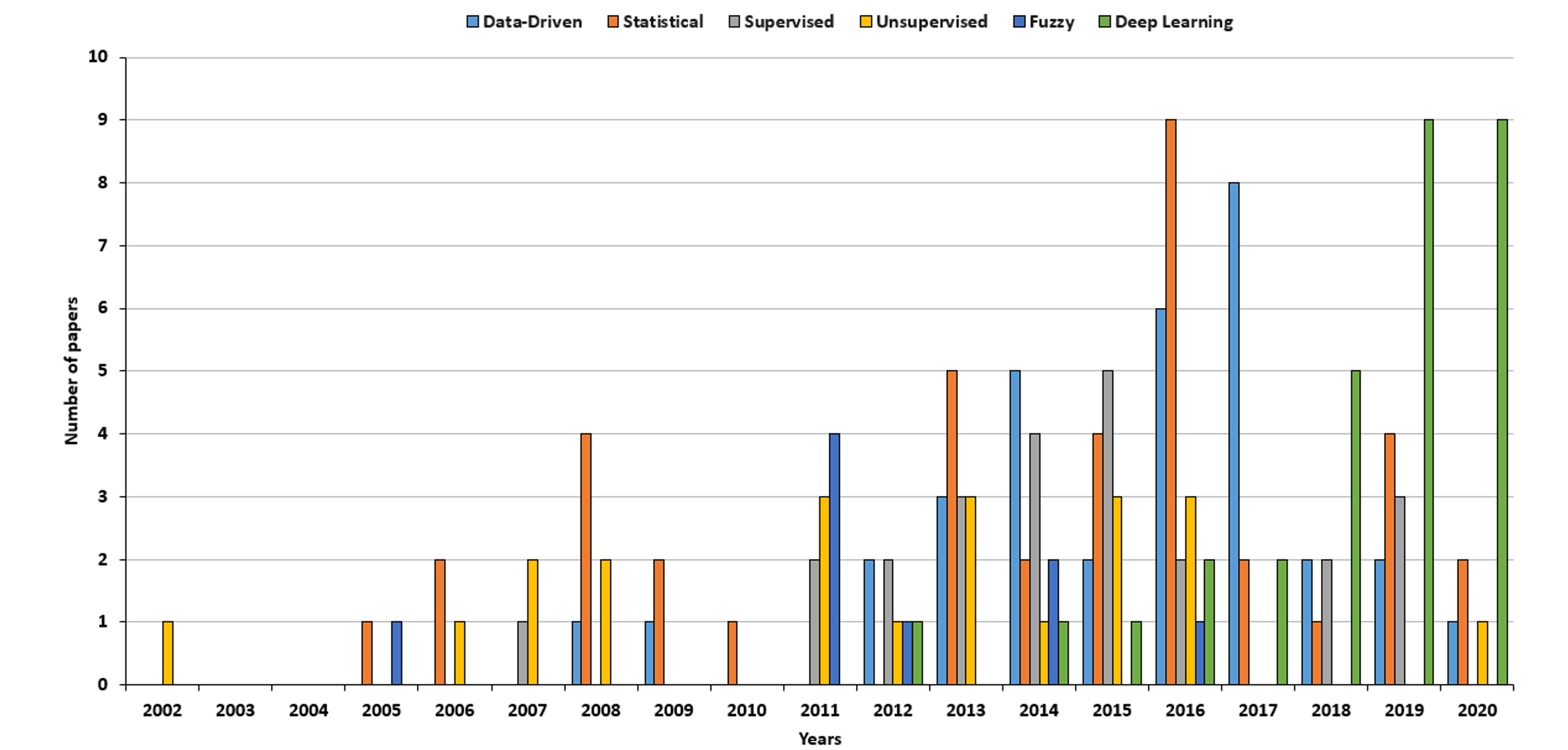}
    \caption{Year-wise distribution of methods}
    \label{fig:yw-num-methods}
\end{figure*}

Further analysis was needed to find the trends and check which methods are going to be adopted in the future. To serve the purpose, we analyzed the number of published papers of each method year-wise. The study shows, that statistical and data-driven methods are used since in the early years; however, there is no single paper on deep learning before 2014. Further analysis shows, that statistical methods in 2016, data-driven in 2017 and deep learning methods in 2019 are widely used for MS segmentation and detection as shown in Figure \ref{fig:yw-num-methods}. In the last two years, the most widely used methods are deep learning models. As the medical imaging data increased, the deep learning methods are adopted and achieved state-of-the-art results.

In future prospective, Transfer learning is one of the domains researchers are currently considering. Currently, researcher use imageNet pre-trained models that are trained on millions of non-medical images and still perform better than most of the traditional learning algorithms. It would be worthy to train a deep learning models with millions of medical images that may perform better than non-medical imaging trained models.

Table \ref{tab:table-img-seq} shows five types of methods that use different MRI modularities; however, most of the studies used FLAIR to diagnose MS lesions. One of the main characteristics of FLAIR is that it suppresses the bright signal of CSF which allows better detection of Hyperintense Lesions (HL). As the name suggests, HL refers to those high-intensity areas in the MRIs which are reflected as lesions in the human brain via demyelination and axonal loss. The other reason is that these images show immense contrast in WML, GML, and other surrounding tissues which makes segmentation easier. Moreover, it is the most sensitive technique to detect WML in MS patients \citep{zwanenburg2010a}. Table \ref{tab:table-img-seq} shows that most of the studies focused on MS lesion (MSL) while few segments and classified WML, GML and tissues. The other columns of the tables illustrate the methods, years and image modularities used to segment and detect MS lesions. All the approaches have advantages over other approaches but also disadvantages as illustrated in Table \ref{tab:pro-cons}.

Table \ref{tab:summary} shows a more detailed comparison of the CAD systems. Most of these approaches are automatic which are more popular and do not need any human assistance during detection, while semi-automatic techniques need human assistance. Two types of datasets were used in these studies. Real datasets consist of real images, including public datasets \citep{int_symp_bio_img}, \citep{mendrik2015mrbrains}, \citep{unknown-h} and patient MRIs from hospitals \citep{cabezas2016a}, \citep{egger2017a}, \citep{jerman2016a} with different MS types including RRMS, PPMS and SPMS \citep{karimaghaloo2016a}, \citep{liu2016a}, \citep{roy2015a}. Few studies focuses on follow-up lesion detection in a time frame of weekly, monthly and yearly such as \citep{cabezas2016a}, \citep{shiee2010a}, \citep{lesjak2016a} and \citep{salem-a}. Some studies compare previous MRIs,  detect new lesions and keep track of the progress of new and existing lesions \citep{ganiler2014a}, \citep{eichinger2017a}, \citep{schmidt-a}. 

Table \ref{tab:summary} also shows that a majority of the experiments were conducted via real datasets and very few using a synthetic dataset with the addition to real datasets. Due to the sensitivity of the problem, the research community was more focused on real data which can reflect the results of real-world problems. Although, synthetic data is used to generate different artifacts like noise, contrasts, and compression which make the evaluation more challenging. 

Table \ref{tab:summary} shows that most of the studies used PPV, TPR, FPR, accuracy, sensitivity, specificity and DSC scores performance metrics; these performance metrics are discussed in Section \ref{sec:5}. But for the sake of comparison, it can be discussed here more in-depth. For a binary classification problem, one of the important metrics is a receiver operating characteristic (ROC) curve with an area under the curve (AUC) value that shows the diagnostic capability of a binary classifier. Performance evaluation can also be rated by professionals as adopted in a study \citep{bonanno2021multiple}. The results were evaluated by two experts who calculated the overall scores. Some studies use different measures for each lesion type like small and moderate \citep{liu2009b} or low and high lesions \citep{sajja2006a}.

\section{Conclusion}

Multiple Sclerosis is a chronic disease that affects millions of people around the world. It affects the central nervous system which leads to severe body damages. To help concerned physicians, automatic diagnosis systems were developed over time to improve the decision-making process. 
In this study, we analyzed the various methods available for detecting and segmenting MS lesions in brain MRIs. The study showed that FLAIR was the most used image among MRI modularities. To evaluate the methods, we used various metrics such as accuracy, sensitivity, specificity, and true positive rate and true negative rate for classification and dice similarity metric for segmentation. 
Our review reveals that there have been 156 papers related to MS lesions segmentation published in various conferences, journals and proceedings during the last two decades. The journal publications represent around 59\%, conferences represent 36\% and the rest 5\% are lecture notes, book chapters, or reports of the total papers. The statistical methods were the most widely used while the fuzzy methods were the least used methods for MS lesion segmentation and detection. The study also found that after 2016, more conference papers were published than journal papers annually. The analysis showed that most papers adopted the statistical methods and showed a tremendous increase in the deep learning papers for MS lesion detection since 2012. The last couple of years also shows that deep learning is gaining more popularity in MS lesion segmentation and detection. 

In future, deep learning methods are most likely to use the transfer learning models because the classic deep learning model needs more resources and time to train the new models from scratch. One of the challenging task is medical data, due to medical sensitive issues, very few samples are publicly available. With such small amount of data, training a deep learning models is a challenging task, deep learning models are data hungry and need tons of data to outperform the traditional algorithms. Therefore, transfer learning would be a better options for MS lesions detection models. Generative adversarial networks \citep{creswell2018generative} can be used to general more medical data for deep learning models. Till now researchers have explored most of the medical imaging data, the graph data and graphical models \citep{wu2019graph} can be employed to reduce the model complexity, improve training time and fast execution. The machine and deep learning based research has wide scope in the area of medical field, however, it still require cooperation between, mathematics, computer science and medical fields.


\small
\clearpage
\onecolumn

\begin{longtable}[c]{|p{1.8cm}|p{0.6cm}|p{1.75cm}|p{1.75cm}|p{2.5cm}|p{1.3cm}|p{1.75cm}|p{3.05cm}|}

\hline

\label{tab:summary}
\multirow{2}{*}{\textbf{Study}} &
  \multirow{2}{*}{\textbf{Year}} &
  \multirow{2}{*}{\textbf{Type}} &
  \multirow{2}{*}{\textbf{Methods}} &
  \multicolumn{2}{c|}{\textbf{Datasets}} &
  \multirow{2}{*}{\textbf{Approach}} &
  \multirow{2}{*}{\textbf{Performance Metric}} \\ \cline{5-6}
 &
  &
  &
  &
  \textbf{Real} &
  \multicolumn{1}{c|}{\textbf{Synthetic}} &
  &
  \\ \hline
\endfirsthead
\multicolumn{8}{c}%
{{\bfseries Table \thetable\ continued from previous page}} \\
\hline
\multirow{2}{*}{\textbf{Study}} &
  \multirow{2}{*}{\textbf{Year}} &
  \multirow{2}{*}{\textbf{Type}} &
  \multirow{2}{*}{\textbf{Methods}} &
  \multicolumn{2}{c|}{\textbf{Datasets}} &
  \multirow{2}{*}{\textbf{Approach}} &
  \multirow{2}{*}{\textbf{Performance Metric}} \\ \cline{5-6}
 &
  &
  &
  &
  \textbf{Real} &
  \multicolumn{1}{c|}{\textbf{Synthetic}} &
  &
  \\ \hline
\endhead
Zeng et al \citep{zeng2013a} &
  2013 &
  Data-driven &
  3D Alpha Matting &
  MICCAI 2008 &
   &
   &
  TPR=0.48, 
  
  PPV=0.59,
  
  DSC=0.52 \\ \hline
Zhong et al \citep{zhong2014a} &
  2014 &
  Data-driven &
  Thresholding + Region Growing &
  26 Patients (RRMS) (Wayne state university) &
   &
  Automatic &
  SI=0.77, 
  
  Correlation=0.96 \\ \hline
Ganiler at el. \citep{ganiler2014a} &
  2014 &
  Data-driven &
  Thresholding &
  20 Patients (10 with 12 month follow-up, 10 with 48 month follow-up) &
   &
   &
  For 12 Months: 
  
  Sens=0.83, 
  
  FDR=0.14,
  
  DSC=0.85 
  
  For 48 Months: 
  
  Sens=0.77,
  
  FDR=0.18,
  
  DSC=0.80 \\ \hline
Storelli et al \citep{storelli2016b} &
  2016 &
  Data-driven &
  Region Growing &
  30 Patients &
   &
  Semi-Automatic &
  DSC=0.78,
  
  TPF=0.81,
  
  FPF=0.14, 
  
  FNF=0.20 \\ \hline
Storelli et al \citep{storelli2016a} &
  2016 &
  Data-driven &
  Region Growing &
  52 Patients (Amsterdam, Graz, London, Milan, Naples and Siana) &
   &
   &
  DSC=0.62,
  
  TPF=0.62,
  
  RMSR =2ml \\ \hline
Mechrez et al \citep{mechrez-a} &
  2016 &
  Data-driven &
  Patch Based + Spatial Consistency &
  MICCAI 2008 &
   &
   &
  Avg: 
  
  TPR=40\%,
  
  PPV=29\%.
  
  DSC=32\% \\ \hline
Cabezas et al \citep{cabezas2016a} &
  2016 &
  Data-driven &
  Image Registration and Deformation &
  36 Patient data (Vall dHebron Hospitals) &
   &
  Automatic &
  DSC=0.86,
  
  FPD=17.8\%,
  
  TPD=70.9\% \\ \hline
Roy et al \citep{roy2017a} &
  2017 &
  Data-driven &
  Adoptive Thresolding &
  The Whole Brain  \citep{unknown-i} &
   &
   &
  JI=90.43\%,
  
  KI=94.88\%,
  
  Acc=92.60\% \\ \hline
Egger et al \citep{egger2017a} &
  2017 &
  Data-driven &
  LGA &
  50 Patients (Brain and Mind Center, University of Sydney, Australia) &
   &
   &
  ICC=0.97, 
  
  DC=0.66 \\ \hline
Eichinger et al \citep{eichinger2017a} &
  2017 &
  Data-driven &
  DIR Subtraction Maps &
  106 Patients &
   &
  Automatic &
  Acc=93\%, 
  
  p=0.013 \\ \hline
Schmidt et al \citep{schmidt-a} &
  2017 &
  Data-driven &
  Image Registration and Fusion &
  20 Patients &
   &
   &
  Sens= 100,
  
  PPV = 83.3\% - 90.1\%
  
  Spec= 88.2\%
  
  NPV=100\% \\ \hline
Stamile et al \citep{stamile2017a} &
  2017 &
  Data-driven &
  Genetic Algorithm &
  5 Patients (RRMS) &
   &
   &
  F-Measure \textgreater 60\% \\ \hline
Zhao et al \citep{zhao2018a} &
  2018 &
  Data-driven &
  Phase Intensities &
  MICCAI 2008 &
   &
  Semi-Automatic &
  Auto DSC = 55\% \\ \hline
Ghribi et al \citep{ghribi2018a} &
  2018 &
  Data-driven &
  Features: 
  
  (GLCM), (GLRLM), 
  
  Volumetric Shape Matrix &
  CHU-HB, MICCAI 2008, MSC 2015 &
   &
  Semi-Automatic &
  Dice = 0.66  $\pm$  0.07
  
  TPR = 0.70  $\pm$  0.12
  
  PPV = 0.67  $\pm$  0.03 \\ \hline
González-Villàa et al \citep{gonzalez2019brain} &
  2019 &
  Data driven &
  Patch Intensity-based multi-atlas segmentation &
  MICCAI 2008 , MICCAI 2016, ISBI 2015 and 2 private datasets &
   &
  Semi-Automatic &

  - DSC for Non-local
  
  Spatial STAPLE 
  
  (NLSS) = 1.96\% 
  
  - DSC for Non-local
  
  Spatial STAPLE
  
  (NLSS) = 2.06\%
  
  improved \\ \hline
Schmidt et al \citep{schmidt2019a} &
  2019 &
  Data driven &
  Intensity based segmentation using time points &
  5 patients from each of three sites: 
  
  Ruhr University of Bochum, 
  
  Johannes Gutenberg University
  
  Mainz, and TUM.
  
  External Validation:
  
  40 patients (MPIP and TUM) &
   &
  Automatic &
  Voxel-wise DC= 0.7 
  
  Lesion-wise detection rate = 0.8, 
  
  False-discovery rate = 0.2 \\ \hline
Pelizzar et al \citep{pelizzari2020semi} &
  2020 &
  Data driven &
  intensity thresholding on QSM and SWI imaging &
  38 MS participants &
   &
  Semi-Automatic &
  Sens: 95.6\%
  
  Spec: 92.1\% \\ \hline
Ge et al \citep{ge2019a} &
  2019 &
  Data-driven &
  Joint Constraints of Low-Rank Representation and Sparse Representation &
  MICCAI BraTS Challenge 2012  
  
  ACCORDION MIND database &
   &
  Automatic 
   &
 
  Acc = 0.7799 \\ \hline
Sajja et al  \citep{sajja2006a} &
  2006 &
  Statistical &
  HMRF-EM &
  23 Patients &
   &
   &
  Low Lesion:
  
  Acc=80.0\% 
  
  High Lesion:
  
  Acc=93.0\% \\ \hline
Garcia et al \citep{garc2008a} &
  2008 &
  Statistical &
  GMM + TLE &
  MICCAI 2008 &
   &
   &
  PPV=69.98\%,
  
  Spec=99.5\% \\ \hline
Dufresne et al \citep{dufresne2020joint} &
  2020 &
  Statistical &
  Optimization framework &
  20 patient \citep{lesjak2016a} &
  BrainWeb &
  &
  BrainWeb: 
  
  Voxel-wise DSC= 0.79
  
  20 patient dataset:
  
  DSC: 0.52 \\ \hline
Shen et al \citep{shan2008a} &
  2008 &
  Unsupervised &
  FCM &
  15 Patients &
  19 Scans from MRico \citep{unknown-j} &
  &
  SI=0.97,
  
  Spec=99.0\%,
  
  Sens=96.0\% \\ \hline
Freifeld et al \citep{freifeld2009a} &
  2009 &
  Statistical &
  CGMM-CE &
  MS Center at Sheba Medical Center, Isreal &
  BrainWeb &
  &
  Avg DI=0.78 \\ \hline
Lie et al \citep{liu2009b} &
  2009 &
  Statistical &
  EM &
  16 Scans (University of Kentucky Hospital) &
   &
  Automatic &
  Small lesion:
  
  DC = 0.63,
  
  Spec=99.9\%, 
  
  Sens=60.03\%
  
  Moderate lesion: 
  
  DC=0.84,
  
  Spec=99.9\%, 
  
  Sens=84.0\% \\ \hline
Karimaghaloo et al \citep{karimaghaloo2013a} &
  2013 &
  Statistical &
  CRF &
  120 Patients \& MICCAI 2013 &
   &
   &
  Avg FPs = 0.60,
  
  Avg PPV = 75\%,
  
  Acc = 93.0\% \\ \hline
Bilello et al \citep{bilello2013a} &
  2013 &
  Statistical &
  Linear Regression + Thresholding &
  98 Scans &
   &
   &
  Sens = 0.87,
  
  Spec =0.91,
  
  PPV=0.87,
  
  NPV=0.91,
  
  Acc = 90\%,
  
  AUC=0.8 \\ \hline
Spies et al \citep{spies2013a} &
  2013 &
  Statistical &
  Probabilistic Maps &
  10 Patients &
  Brain-Web &
  Automatic &
  DC=0.6, 
  
  Acc = 97.0\% \\ \hline
Gao et al \citep{gao2014a} &
  2014 &
  Statistical &
  Bias Field Estimation &
  MICCAI 2008 &
   &
   &
  Rater 1: DSC=0.58,
  
  Avg Spec=0.98,
  
  Avg FNR=0.1368
  
  Rater 2: DSC=0.54,
  
  Avg Spec=0.933,
  
  Avg FNR=0.1431 \\ \hline
Puonti et al \citep{puonti2016a} &
  2016 &
  Statistical &
  Contrast-Adaptive Probabilistic Models &
  MICCAI 2008 &
   &
   &
  TPR=0.41,
  
  PPV=0.40 \\ \hline
Karimaghaloo et al \citep{karimaghaloo2015a} &
  2015 &
  Statistical &
  THAT-CRF &
  120 Scans &
   &
   &
  FDR = 0.196,
  
  Sens = 94.8\% \\ \hline
Jain et al \citep{jain2016a} &
  2016 &
  Statistical &
  EM &
  12 Patients and 10 Patients scans &
   &
   &
  Median : DS = 0.63,
  
  PPC = 0.96 \\ \hline
Liu et al \citep{liu2016a} &
  2016 &
  Statistical &
  SPREAD &
  7 Patients (RRMS) &
   &
   &
  TPR=73.98\% \\ \hline
Karimaghaloo et al \citep{karimaghaloo2016a} &
  2016 &
  Statistical &
  multi-level AMCRF &
  2270 scans \& 120 scans (RRMS) &
   &
   &
  For Scans 2270: 
  
  Sens=90\%,
  
  Avg FPC=0.17 
  
  For Scans 120: 
  
  Sens=91\%, 
  
  Avg FPC=0.5,
  
  FDR=0.33 \\ \hline
Strumia et al \citep{strumia2016a} &
  2016 &
  Statistical &
  Geometric Brain Model &
  MICCAI 2008 &
   &
   &
  VD=36.5\%,
  
  TPR=70.4,
  
  SD=2.8 \\ \hline
Valverde et al \citep{valverde2017a} &
  2017 &
  Statistical &
  probabilistic prior maps &
  MS-BrainS13 \citep{mendrik2015mrbrains} &
   &
   &
  \begin{small}AVD-CSF=0.4$\pm$0.07\%,
  
  AVD-GM=0.8$\pm$0.12 \%,
  
  AVD-WM= 0.11$\pm $0.16\% \end{small}\\ \hline
Dworkin et al \citep{dworkin2018a} &
  2018 &
  Statistical &
  Probability estimation &
  60 subjects - National Institute of Neurological Disorders and Stroke in Bethesda, Maryland. &
   &
  Automatic &
  Correlation = 0.97 
  
  and p \textless 0.001 \\ \hline
Ghribi et al \citep{ghribi2019a} &
  2019 &
  Statistical &
  Gaussian Mixture model &
  UHST, MICCAI 2008, Longitudinal database, ISBI 2015 &
   &
  Automatic &
  
  
  Avg Acc = 0.75 $\pm$ 0.12 
  
  DSC = 0.78 $\pm$ 0.04, 
  
  Sens = 0.85 $\pm$ 0.04, 
  
  Spec = 0.89  $\pm$  0.04 \\ \hline
Pota et al \citep{pota2019a} &
  2019 &
  Statistical &
  Multivariate analysis &
  81 patients from Centre of the “Federico II” University Hospital (Naples, Italy). &
   &
  Automatic &
  Acc = 89.1\% 
  
  Sens = 93.8\% 
  
  NPV = 81.5\%
  
  PPV = 91.6\% \\ \hline
Freire and Ferrari \citep{freire2019a} &
  2019 &
  Statistical &
  White matter mask estimation &
  ISBI2015 &
   &
  Automatic &
  Sens =  99\%,
  
  Spec =  98\% \\ \hline
Wang et al \citep{wang2019a} &
  2019 &
  Statistical &
  MAMC &
  MICCAI2008 &
   &
  Automatic &
  
  Mean TPR = 0.34
  
  Mean PPV =  0.45
  
  Mean DSC = 0.38 \\ \hline
Wang et al \citep{wang2020a} &
  2020 &
  Statistical &
  Sparse Bayesian model &
  MICCAI 2008 &
   &
  Automatic &
  Mean TPR = 0.57,
  
  Mean PPV = 0.38,
  
  Mean DSC = 0.42 \\ \hline

Roy et al \citep{roy2013a} &
  2013 &
  Supervised &
  SVM &
  MICCAI 2008 &
   &
  Automatic &
  F1-Score = 0.5 \\ \hline
Weiss et al \citep{weiss2013a} &
  2013 &
  Supervised &
  Dictionary Learning &
  MICCAI 2008 &
  BrainWeb &
  &
  Mean SD: 
  
  TPR = 33\%,
  
  PPV = 37\% \\ \hline
Loizou et al \citep{loizou2011b} &
  2013 &
  Supervised &
  SVM &
  38 Patients &
   &
  Automatic &
  
  Acc = 69 $\pm$ 3.1,
  
  Sens = 74 $\pm$ 7.3,
  
  Spec = 65 $\pm$ 10,
  
  FNR = 26 $\pm$ 7.3,
  
  FPR = 35 $\pm$ 10 \\ \hline
Roy et al \citep{roy2015a} &
  2015 &
  Supervised &
  patch-based segmentation &
  10 Patients (8 RRMS, 1 SPMS, 1 PPMS) &
   &
   &
  Median:
  
  TPR = 0.532
  
  FPR = 0.432
  
  Sens = 57.5\% \\ \hline
Gomez et al \citep{g2015a} &
  2016 &
  Supervised &
  Random Forest &
  13 Patients &
   &
  Automatic &
  Avg DS=0.58 \\ \hline
Ghafoorian et al \citep{ghafoorian2016a} &
  2016 &
  Supervised &
  AdaBoost &
  362 Patients \citep{norden2011a} &
   &
  Automatic &
  TPR=80.0\% \\ \hline
Santos et al \citep{santos2016a} &
  2016 &
  Supervised &
  Multi-Layer Perceptron Network &
  MICCAI 2008 &
   &
   &
  3D-Score = 83.26\% \\ \hline
Salem et al \citep{salem-a} &
  2018 &
  Supervised &
  Logistic regression &
  60 patients &
   &
  Automatic &
  DSC= 0.77,
  
  TPR = 74.30\%,
  
  FP=11.86\% \\ \hline
Wang et al \citep{wang-a} &
  2018 &
  Supervised &
  Fourier entropy and Jaya algorithm &
  38 patients - eHealth laboratory &
   &
  Automatic &
  Sens = 97.40 $\pm$ 0.60\%
  
  Spec = 97.39 $\pm$ 0.65\%
  
  Acc  = 97.39 $\pm$ 0.59\% \\ \hline
Han and Hou \citep{han2019a} &
  2019 &
  Supervised (Chapter) &
  feed forward neural network + Genetic algorithm &
  eHealth laboratory - 38 subjects (MS) and 26 healthy control &
   &
  Automatic &
  Acc = 91.95  $\pm$  1.19 \\ \hline
Zhang et al \citep{zhang2019a} &
  2019 &
  Supervised &
  Random Forest &
  84 patients (private) &
   &
  Semi-Automatic &
  Avg : 
  
  Acc= 0.82 
  
  Sens= 0.95 
  
  Spec= 0.33 
  
  PPV=  0.84 
  
  Balanced Acc= 0.64 
  
  DOR= 10.50 
  \\ \hline

Jerman et al \citep{jerman2016a} &
  2016 &
  Unsupervised &
  Unsupervised Mixture model + Decision Trees &
  18 Patients (University medical center Ljubljana) &
   &
   &
  DS = 0.73,
  
  Sens = 0.90,
  
  PPV = 0.61 \\ \hline

Steenwijk et al \citep{steenwijk2013a} &
  2013 &
  Unsupervised &
  KNN + TTPs &
  20 Patients &
   &
   &
  ICC = 0.93,
   
  SI = 0.75 $\pm$ 0.08 \\ \hline
Lyksborg et al \citep{lyksborg2012a} &
  2012 &
  Unsupervised &
  KNN + Markov Random Field &
  71 Images &
   &
   &
  Median: 
  
  SI = 0.702, 
  
  OF = 0.743 \\ \hline
Bonanno et al \citep{bonanno2021multiple} &
  2020 &
  Unsupervised &
  watershed for segmentation and cluster analysis &
  20 MRI images &
   &
  Automatic &
  Spec: 87\% 
  
  AUC: 77\% \\ \hline
Admiraal-Behloul et al \citep{admiraal-behloul2005a} &
  2005 &
  Fuzzy &
  Fuzzy Inference &
  100 Scans &
   &
  Automatic &
  Correlation = 0.98
  
  SI = 0.75 \\ \hline
Yoo et al \citep{yoo2014a} &
  2014 &
  Deep learning &
  Deep Belief Network + Random Forest &
  581 Patients (1450 T2 and PD scans) &
   &
  Automatic &
  Avg: TPR = 58  $\pm$  17
  
  PPV=35 $\pm$ 24
  
  DSC = 38 $\pm$ 19 \\ \hline
Brosch et al \citep{brosch2015a} &
  2015 &
  Deep learning &
  Deep Convolutional Encoder Networks &
  500 Scan in house data \& MS lesion Segmentation Challenge 2008 &
   &
  Automatic &
  TPR = 39.71\%
  
  PPV = 41.38\%
  
  DSC = 35.32\% \\ \hline
Brosch et al \citep{brosch2016a} &
  2016 &
  Deep learning &
  CNN &
  MICCAI 2008 and ISBI 2015 &
   &
  Automatic &
  DSC=63.83\%,
  
  TPR=52.49\%,
  
  FPR=36.10\%,
  
  VD=32.89\% \\ \hline
Birenbaum et al \citep{birenbaum2016a} &
  2016 &
  Deep learning &
  CNN &
  14 Patients (ISBI 2015) \citep{int_symp_bio_img} &
   &
   &
  DS=0.627 \\ \hline
Prieto et al \citep{prieto-a} &
  2017 &
  Deep learning &
  CNN &
  900 Subjects (CLIMB) \citep{unknown-h} &
   &
  Automatic &
  Acc = 84.0\% \\ \hline
Valverde et al \citep{valverde2017b} &
  2017 &
  Deep learning &
  CNN &
  MICCAI 2008 and 2 Private datasets &
   &
  Automatic &
  Rater 1: 
  
  VD=62.5\%
  
  TPR=55.5\%
  
  FPR=46.88\% 
  
  Rater 2: 
  
  VD=40.8\%
  
  TPR=68.7\%
  
  FPR=46.0\%
  
  Overall Score = 87.12 \\ \hline
Zhang et al \citep{zhang2018a} &
  2018 &
  Deep learning &
  CNN &
  eHealth laboratory (38 individuals eHeath dataset), and Private (26 individuals) &
   &
  Automatic &
  Sens = 98.22\%,
  
  Spec = 98.24\%,
  
  Acc =  98.23\% \\ \hline
Roya et al \citep{roy2018a} &
  2018 &
  Deep learning &
  CNN &
  100 MS patients, ISBI 2015 Challenge &
   &
  Automatic &
  100 patient data: 
  
  DSC=0.5639
  
  FPR = 0.3077
  
  PPV = 0.6040
  
  ISBI dataset:
  
  DSC = 05243
  
  FPR = 0.1103
  
  PPV=0.8660
  
  Score= 90.48 \\ \hline
Wang et al \citep{wang-b} &
  2018 &
  Deep learning &
  CNN &
  38 patient – eHealth and26 patient - private dataset
  
  Take dataset reference from table 1 in same paper &
   &
  Automatic &
  Sens = 98.77  $\pm$  0.35\%,
  
  Spec= 98.76 $\pm$ 0.58\%,
  
  Acc = 98.77  $\pm$  0.39\% \\ \hline
Tanya et al \citep{nair2020a} &
  2020 &
  Deep learning &
  3D-CNN &
  Private Dataset of 1064 patients &
   &
  Automatic &
  Voxel-wise: 
  
  
  ROC = 99.78
  
  Lesion-wise:
  
  
  ROC= 100 \\ \hline
Zhang et al \citep{zhang2018b} &
  2018 &
  Deep learning &
  MS-GAN &
  126 scans – private dataset &
   &
  Automatic &
  DSC = 0.672
  
  Recall = 0.692
  
  Prec = 0.724
  
  F-score = 0.708 \\ \hline
Valverdea et al \citep{valverde2019a} &
  2019 &
  Deep learning &
  CNN &
  60 patients (Hospital Vall d’Hebron, Barcelona, Spain), ISBI 2015 &
   &
  Automatic &
  60 patients dataset:
  
  DSC = 0.52 $\pm$ 0.16,
  
  Sen = 0.60 $\pm$ 0.21,
  
  Prec = 0.75 $\pm$ 0.21
  
  ISBI dataset:
  
  DSC = 0.58 $\pm$ 0.16,
  
  Sen = 0.48 $\pm$ 0.19,
  
  Prec=0.84 $\pm$ 0.13
  
  Score = 90.32 \\ \hline
Aslani et al \citep{aslani2019a} &
  2019 &
  Deep learning &
  CNN &
  NRU dataset (Ospedale San Raffaele, Milan, Italy, Consists of 37 MS patients ISBI dataset &
   &
  Automatic &
  
  
  DSC = 0.7067 
  
  PPV =  0.6844
  
  Lesion TPR = 0.6136
  
  Lesion FPR = 0.1284 
  
  VD=0.1488 
  
  SrD=1.577
  
  HD = 8.368 \\ \hline
Salem et al \citep{salem-b} &
  2019 &
  Deep learning &
  Encoder-Decoder U-net &
  Private and ISBI 2015 &
   &
  Automatic &
  DSC = 0.63  $\pm$  0.13
  
  Sens = 0.55 $\pm$ 0.16
  
  Prec = 0.79 $\pm$ 0.14
  
  Score = 91.33 \\ \hline
Hashemi et al \citep{hashemi-a} &
  2019 &
  Deep learning &
  3D- fully Connected-Dense Network &
  MSSEG 2016 and ISBI &
   &
  Automatic &
  MSSEG dataset:
  
  DSC = 69.9\% 
  
  ISBI dataset:
  
  DSC = 65.74\% \\ \hline
Gabr et al \citep{gabr2019a} &
  2019 &
  Deep learning &
  fully convolutionalneural network (FCNN) &
  CombiRx clinical trial - (a private dataset) 1008 patients with RRMS &
   &
  Automatic &
  Mean (95\% CI) 
  DSC for:  
  
  WM:
  
  0.95 (0.92-0.98),
  
  GM:
  
  0.96 (0.93-0.98),
  
  CSF:
  
  0.99(0.98-0.99)
  \\ \hline
Feng et al \citep{feng2019a} &
  2019 &
  Deep learning &
  FCNN &
  ISBI 2015 &
   &
  Automatic &
  DSC = 0.684 
  \\ \hline
Narayana et al \citep{narayana2018a} &
  2018 &
  Deep learning &
  CNN &
  CombiRx database &
   &
  Automatic &
  WM and GM : 
  
  DSC = 094 
  
  CF:
  
  DSC = 0.94 
  
  Hyperintense lesion:
  
  DSC = 0.85 \\ \hline
Rosa et al \citep{rosa2019a} &
  2019 &
  Deep learning &
  Shallow and Deep learning Architectures – K-NN and CNNs &
  73 patients, 50 females and 23 males &
   &
  Automatic &
  DC = 60\%, 
  
  lesion-wise TPR=69\%,
  
  FPR =26\% \\ \hline
Li et al \citep{li2019a} &
  2019 &
  Deep learning &
  multi-scale convolutional-stack aggregation model &
  MICCAI WMH Segmentation Challenge 2017 – 60 cases, private 30 cases. (hospitalin Munich) &
   &
  Automatic &
  MICCAI WMH :
  
  DSC = 80.09\%
  
  Recall = 86.96\%
  
  F1-score=76.73\%
  
  MS Dataset:
  
  DSC = 76.93\%
  
  Recall = 93.16\%
  
  F1-score=79.57\% \\ \hline
Aslani et al \citep{aslani2019b} &
  2019 &
  Deep learning &
  CNN &
  ISBI 2015 &
   &
  Automatic &
  Rater 1: 
  
  DSC = 0.6940 
  
  Lesion TPR = 0.7840 
  
  Lesion FRP = .4970 
  
  Rater 2: 
  
  DSC = 0.6640
  
  Lesion LPR=0.6950 
  
  Lesion FPR = 0.4420 \\ \hline
Sepahvand et al \citep{sepahvand2019a} &
  2019 &
  Deep learning &
  3D-CNN &
  1068 patient’s dataset – a private &
   &
  Automatic &
  Sens =  80.11\%,
  
  Spec = 79.16\%,
  
  Prec = 91.82\%
  
  Acc = 80.21\% \\ \hline
Mckinley et al \citep{mckinley2020a} &
  2020 &
  Deep learning &
  DeepSCAN MS &
  Three private datasets:
  
  - Zurich Dataset (Radiology CenterBethanien)
  
  - Munich dataset (Radiology CenterBethanien)
  
  - Bern dataset (MS cohort databank of the University of Bern) &
   &
  Automatic &
  Zurich Dataset
  
  Acc = 75\%,
  
  Sens = 0.60,
  
  PPV = 0.84
  
  Munich dataset
  
  Acc = 85\%,
  
  Sens = 0.72,
  
  PPV = 1.00
  
  Bern dataset
  
  Acc = 91\%,
  
  Sens = 1.00,
  
  PPV = 0.59 \\ \hline
Salem et al \citep{salem2020a} &
  2020 &
  Deep learning &
  FCNN &
  60 patients &
   &
  Automatic &
  DSC = 0.83,
  
  TPR = 83.09\%,
  
  FPR = 9.36\% \\ \hline
Sepahvand et al \citep{sepahvand2020cnn} &
  2020 &
  Deep learning &
  U-Net and CNN &
  1677 patient brain images &
   &
  Automatic &
  Sens= 0.69, 
  
  Spec= 0.97
  
  AUC=0.95
  
  ROC curves:
  
  voxel level= 0.9083 $\pm$ 0.0041
  
  lesion level= 0.9466 $\pm$ 0.0035 \\ \hline
Ye et al \citep{ye2020deep} &
  2020 &
  Deep learning &
  DBSI with Deep Neural Networks &
  38 patients &
   &
  Automatic &
  Concordance: 93.4\% \\ \hline
Gessert et al \citep{gessert2020multiple} &
  2020 &
  Deep learning &
  fully-convolutional CNN based on U-Net &
  ZURICH and 97 labeled pairs from 33 patients &
   &
   &
   Dice=58.3 
   
   TPR=63.7
   
   FPR=28.3 \\ \hline
La Rosa et al \citep{la2020multiple} &
  2020 &
  Deep learning &
  Encoder-Decorder-FCNN and 3D-U-net &
  Basel + Lausanne datasets of 90 patients &
   &
  Automatic 
  &
  Cross Evaluation:
  
  Detection Dice=0.6
  
  AVD= 0.13
  
  PPV= 0.64
  
  LTPR= 0.69
  
  LFPR= 0.27 \\ \hline
Krüger et al \citep{kruger2020fully} &
  2020 &
  Deep learning &
  U-net 3D-CNN &
  Collected routine data for training and  Zurich and  Dresden for evaluation &
   &
  Automatic &
  Zurich : 
  
  Sen=0.54
  
  Dice=0.39
  
  Fp count=0.63
  
  FPR=0.42 
  
  Dresden: 
  
  Sen=0.67
  
  Dice=0.52
  
  FP count=0.34
  
  FPR=0.4 \\ \hline
\end{longtable}

\clearpage
\twocolumn
\normalsize


%
%


\setcitestyle{number}
\bibliographystyle{unsrt}
\bibliography{Manuscript}
\end{sloppypar}
\end{document}